\documentclass[aps,prl,times,twocolumn,amsmath,amssymb,superscriptaddress,floatfix]{revtex4}

\usepackage{color}
\newcommand{\red}{\color{red}}
\usepackage{graphicx}
\usepackage{subfigure}
\usepackage{bbold}
\usepackage{verbatim}
\usepackage{float}
\usepackage{enumerate}
\usepackage{amsfonts}
\usepackage{multirow}
\usepackage[colorlinks,bookmarks=false,citecolor=blue,linkcolor=red,urlcolor=blue]{hyperref}

\newcommand{\dg}{^\dagger}
\newcommand{\pdg}{^{\phantom\dagger}}

\begin{document}


\title{
Giant magnetoelastic-coupling driven spin-lattice liquid state in molybdate pyrochlores
}


\author{Andrew Smerald}
\email{andrew.smerald@gmail.com}

\affiliation{Max Planck Institut f\"ur Festk\"orperforschung, Heisenbergstra\ss e 1, D-70569 Stuttgart, Germany}

\author{George Jackeli}
\altaffiliation{Also at Andronikashvili Institute of Physics, 0177 Tbilisi,
Georgia}

\affiliation{Max Planck Institut f\"ur Festk\"orperforschung, Heisenbergstra\ss e 1, D-70569 Stuttgart, Germany}
\affiliation{Insitute for Functional Matter and Quantum Technologies, University of Stuttgart, Pfaffenwaldring 57, D-70569 Stuttgart, Germany}


\date{\today}


\begin{abstract}  
We propose the idea of a spin-lattice liquid, in which spin and lattice degrees of freedom are strongly coupled and remain disordered and fluctuating down to low temperatures.
We show that such a state arises naturally from a microscopic analysis of a class of molybdate pyrochlore compounds, and is driven by a giant magnetoelastic effect.
Finally, we argue that this could explain some of the experimental features of  Y$_2$Mo$_2$O$_7$.
\end{abstract}

\maketitle


Frustration is at the heart of the search for unusual states of matter, and can promote spin-liquid behaviour, in which the spins remain disordered and fluctuating down to low temperature \cite{lmm,balents10,savary16}.
However, the interaction of spins with lattice degrees of freedom is typically expected to result in long-range order via an ``order-by-disorder''-type mechanism \cite{villain80,shender82}.
For example, magnetoelastic coupling can drive a structural distortion that selects a small subset of the otherwise extensively degenerate spin configurations \cite{yamashita00,tchernyshyov02a,tchernyshyov02b,penc04,dimatteo04,dimatteo05,bergman06,shannon10}.
Nevertheless, there is the intriguing possibility that spin and lattice degrees of freedom can become tightly coupled, but, rather than ordering, remain fluctuating at low temperature, and we dub the resultant state a spin-lattice liquid.

In order to identify candidate materials for such a spin-lattice liquid state, there are a number of characteristic features that are required.
These include strong magnetoelastic coupling, an absence of spin ordering at temperatures well below the interaction energy and evidence for incoherently disordered lattice degrees of freedom at low temperature.
All of these characteristics have been experimentally observed in the pyrochlore molybdates, {\it R}$_2$Mo$_2$O$_7$ ({\it R}$=$Y,Tb,Dy,Ho,Er,Tm,Yb,Lu \cite{clark13,shinaoka18}), of which Y$_2$Mo$_2$O$_7$ is the best studied example  \cite{greedan86,reimers88,raju92,dunsiger96,gingras97,gardner99,booth00,keren01,sagi05,greedan09,ofer10,silverstein14,thygesen17}.
However, many aspects of their low-temperature behaviour remain mysterious, despite more than three decades of study.

In this Letter we analyse the Mo pyrochlores and show that at intermediate temperatures they may support a state in which spin and lattice degrees of freedom become strongly coupled and explore an extensive manifold of degenerate configurations. 
This state, which we refer to as a spin-lattice liquid, both serves as a first example of a class of states, as well as providing an explanation for the unusual behaviour of the {\it R}$_2$Mo$_2$O$_7$ materials.
We use microscopic analysis to show that spin-orbital and lattice degrees of freedom would individually be expected to show behaviour similar to the classical spin-liquid ``spin ice'' \cite{bramwell01}.
However, due to strong magnetoelastic coupling, a ``spin-lattice'' liquid phase emerges at low temperature.
Since the magnetoelastic coupling plays a decisive and nonperturbative role in selecting the low-energy configurations, we refer to it as a ``giant'' magnetoelastic effect.

\begin{figure}[t]
\centering
\includegraphics[width=0.5\textwidth]{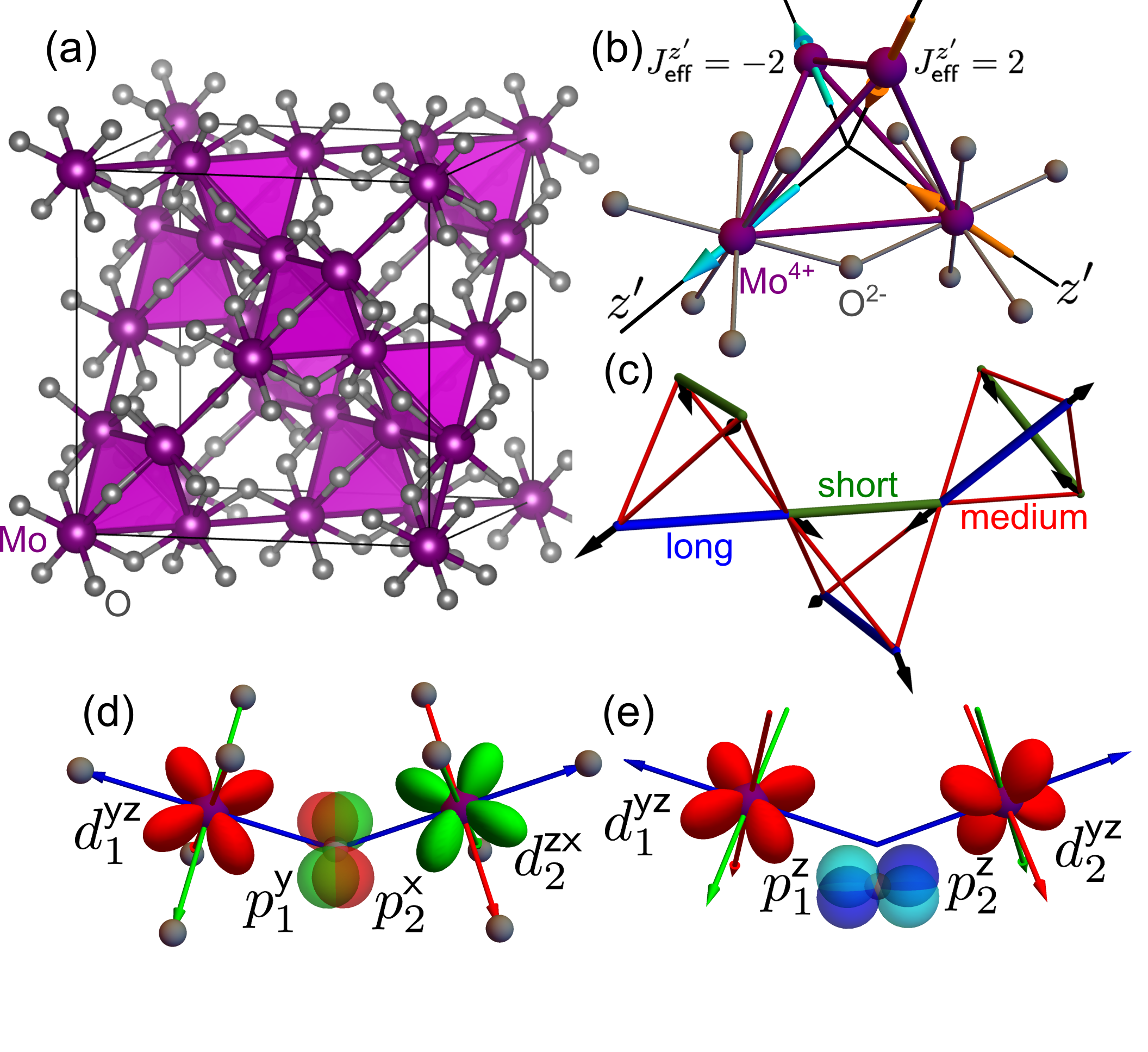}
\caption{\footnotesize{
{\it R}$_2$Mo$_2$O$_7$ pyrochlores.
(a) Partial view of the average {\it Fd$\bar{3}$m} structure of {\it R}$_2$Mo$_2$O$_7$, showing the Mo pyrochlore lattice (purple) and surrounding O octahedra (grey) within a unit cell (black cuboid).
(b) $J^{z^\prime}_{\sf eff} = \pm 2$ states are represented by arrows pointing in/out of the Mo tetrahedron along the $z^\prime$ local axes.
(c) 2-in-2-out lattice displacements (black arrows) create one long (blue), four medium (red) and one short (green) Mo-Mo bond on each tetrahedron.
(d) $\pi$-type $dpd$ hopping path, shown for idealised regular oxygen octahedra.
The path shown is $d_1^{\sf yz}$-$(p_1^{\sf y}$:$p_2^{\sf x})$-$d_2^{\sf yz}$, where Mo sites are labelled 1 and 2 and have associated local coordinates ($x$=red, $y$=green, $z$=blue). 
(e) $\sigma$-type hopping path allowed by trigonal distortion of the oxygen octahedra and exemplified by $d_1^{\sf yz}$-$(p_1^{\sf z}$:$p_2^{\sf z})$-$d_2^{\sf yz}$.
}}
\label{fig:dpdhopping}
\end{figure}

The {\it R}$_2$Mo$_2$O$_7$ materials crystallise into the well-known {\it Fd$\bar{3}$m} space group, where Mo$^{4+}$ ions form a pyrochlore lattice of corner-sharing tetrahedra and are surrounded by oxygen octahedra \cite{reimers88} (see Fig.~\ref{fig:dpdhopping}{\red a},{\red b}).
%
%
At low temperature ($T_{\sf f} \approx 20K$) a spin-glass transition is observed \cite{greedan86,gingras97}, whose origin remains an open question \cite{shinaoka18}.
However, this occurs well below the Curie-Weiss temperature ($\theta_{\sf cw} =-200K$ for Y$_2$Mo$_2$O$_7$ \cite{gardner04}), which approximately measures the strength of magnetic interactions.
As such there is a wide temperature range, $T_{\sf f} < T<\theta_{\sf cw}$ in which spin correlations are expected to be significant, but no spin ordering is observed.

At the same time, extensive studies of the crystal structure have shown that there are significant low-temperature local distortions away from the average {\it Fd$\bar{3}$m} structure that survive up to temperatures in excess of $\theta_{\sf cw}$, but no structural phase transition \cite{booth00,keren01,sagi05,ofer10,greedan09,thygesen17}.
Recent pair-distribution-function analysis has shown that for each tetrahedron two Mo$^{4+}$ ions move towards and two away from the tetrahedral centre, forming a highly-degenerate ``2-in-2-out'' distortion pattern \cite{thygesen17} (see Fig.~\ref{fig:dpdhopping}{\red c}).
There is a concomitant distortion of the O octahedra, such that both the Mo-O bond length and the local crystal field remains almost invariant.
However, the nearest-neighbour Mo-Mo distances and the Mo-O-Mo bond angles change significantly.

Magnetoelastic coupling provides a link between the spin and lattice degrees of freedom, and there is evidence from NMR and $\mu$sr measurements that it is significant in Y$_2$Mo$_2$O$_7$ \cite{keren01,sagi05,ofer10}.
In consequence, the experimental prerequisites appear to be in place to think the {\it R}$_2$Mo$_2$O$_7$ materials may realise a spin-lattice liquid state.
In order to explore whether this is the case, we use microscopic arguments to understand the interplay between spin, orbital and lattice, and derive a simple effective model in an attempt to capture the essence of their low-temperature behaviour.
%


{\it Local electronic states}: The starting point is to consider the local physics of the Mo$^{4+}$ ions.
Mo$^{4+}$ has two electrons in the $4d$-shell, and these are localised by an on-site Coulomb repulsion, $U$, that is large compared to the bandwidth \cite{shinaoka18}.
The local cubic crystal field due to the surrounding O$^{2-}$ octahedron (see Fig.~\ref{fig:dpdhopping}{\red a},{\red b}) splits the $4d$ energy levels into a low-energy $t_{2g}$ manifold of {\sf xy}, {\sf yz} and {\sf zx} orbitals that point between the O ions, and a high-energy $e_{g}$ manifold of orbitals pointing towards O ions \cite{fazekas}.
The $t_{2g}$ manifold is split by Hund's coupling ($J_{\sf H} \approx 0.5$eV \cite{shinaoka13}), which favours parallel spins, and therefore selects a 9-fold degenerate set of low-energy states, labelled by spin $S=1$ and effective orbital angular momentum $L_{\sf eff}=1$ \cite{abragam}.
This is further split by a non-cubic component of the crystal field with trigonal symmetry that results from a compression of O octahedra along the local $ z^\prime$-axes pointing into/out of Mo tetrahedra (see Fig.~\ref{fig:dpdhopping}{\red b}) and spin-orbit coupling.
The characteristic energy scales are repectively $\Delta_{\sf trig}\approx160$meV and $\lambda_{\sf SO} \approx 40$meV \cite{shinaoka13}.
The result is a low-energy doublet, labelled as $J^{z^\prime}_{\sf eff} = \pm2$, where  ${\bf J}_{\sf eff} = {\bf S}+{\bf L}_{\sf eff}$, that is separated from the first excited state by an energy gap $\Delta E = \lambda_{\sf SO}[1 - \lambda_{\sf SO}/\Delta_{\sf trig}+\dots]\approx 30$meV$\approx 300$K  ($k_B=1$) \cite{supmat}.
%


{\it Exchange interaction}: 
An effective low-energy model can be derived by considering superexchange interactions between neighbouring $J^{z^\prime}_{\sf eff} = \pm2$ doublets.
This is different from the approach taken by a number of other theories, which assume an isotropic $S=1$ spin on every Mo site \cite{saunders07,andreanov10,shinaoka14}.
The geometry is simplest in the case of idealised undistorted oxygen octahedra, where local cubic axes can be aligned with the O octahedra, as shown in Fig.~\ref{fig:dpdhopping}{\red d}.
Superexchange occurs via a single intermediate oxygen with Mo-O-Mo bond angle $\alpha = 2 \arctan 2\sqrt{2} \approx 141^\circ$.
Within the local cubic axes the dominant hopping channel is off-diagonal, bond-dependent and follows paths of the type $d_1^{\sf yz}$-$(p_1^{\sf y}$:$p_2^{\sf x})$-$d_2^{\sf zx}$, where $p_1^{\sf y}$ and $p_2^{\sf x}$ denote orbitals on the same oxygen ion but in the local axes of the two different Mo ions (1 and 2 in Fig.~\ref{fig:dpdhopping}{\red d}).
It can be seen that the $d$-$p$ bonds involve a lateral overlap of orbitals, and this is known as $\pi$-bonding \cite{harrison}.

We work in the Mott insulator picture, and for two electrons in the $t_{2g}$ manifold, the derivation of a spin-orbital Hamiltonian from a multiorbital Hubbard model is a standard procedure \cite{supmat,khaliullin01,horsch03,khaliullin05}.
In order to arrive at an effective low-temperature model, it is required to project the full spin-orbital Hamiltonian for an $S=1$, $L_{\sf eff}=1$ configuration onto the $J^{z^\prime}_{\sf eff} = \pm 2$ doublet.
Since no matrix elements connect $J^{z^\prime}_{\sf eff} = 2$ to $J^{z^\prime}_{\sf eff} = - 2$, the effective Hamiltonian takes the form of a classical Ising model $\mathcal{H}_{\sf is} = J_{\sf is} \sum_{\langle ij\rangle} \sigma_i \sigma_j$ \cite{supmat}, where $\sigma = \pm 1$ labels the doublet states, and can be thought of as an Ising ``spin'' pointing into or out of tetrahedra (see Fig.~\ref{fig:dpdhopping}{\red b}).
$J_{\sf is}>0$ favours antiferromagnetic alignment in the local trigonal axes, which corresponds to ferromagnet exchange in crystallographic axes, and low-energy states consist of the extensively-degenerate spin-ice configurations with a 2-in-2-out arrangement.
$J_{\sf is}<0$ favours ferromagnetic alignment in the local trigonal axes, corresponding to antiferromagnet exchange in the crystallographic axes and therefore unfrustrated all in/out arrangements on tetrahedra.
In the case of undistorted cubic octahedra we find $J_{\sf is}<0$ for any value of $J_{\sf H}/U$.

In reality the oxygen octahedra are trigonally distorted, and in Y$_2$Mo$_2$O$_7$ the Mo-O-Mo bond angle in the average {\it Fd$\bar{3}$m} structure is $\alpha = \alpha_{\sf av} = 127^\circ$ \cite{reimers88,greedan09,thygesen17}.
The displacement of the O relative to the undistorted case opens up a $\sigma$-type hopping channel, where orbitals directly overlap one another \cite{harrison}, and this is exemplified by the path $d_1^{\sf yz}$-$(p_1^{\sf z}$:$p_2^{\sf z})$-$d_2^{\sf yz}$ shown in Fig.~\ref{fig:dpdhopping}{\red e}. 
The $\sigma$ channel becomes comparable to the original $\pi$-bonding channel for a relatively small trigonal distortion, and as a result $J_{\sf is}>0$ for $\alpha = 127^\circ$, favouring a 2-in-2-out set of low-energy configurations \cite{supmat}.

\begin{figure}[t]
\centering
\includegraphics[width=0.49\textwidth]{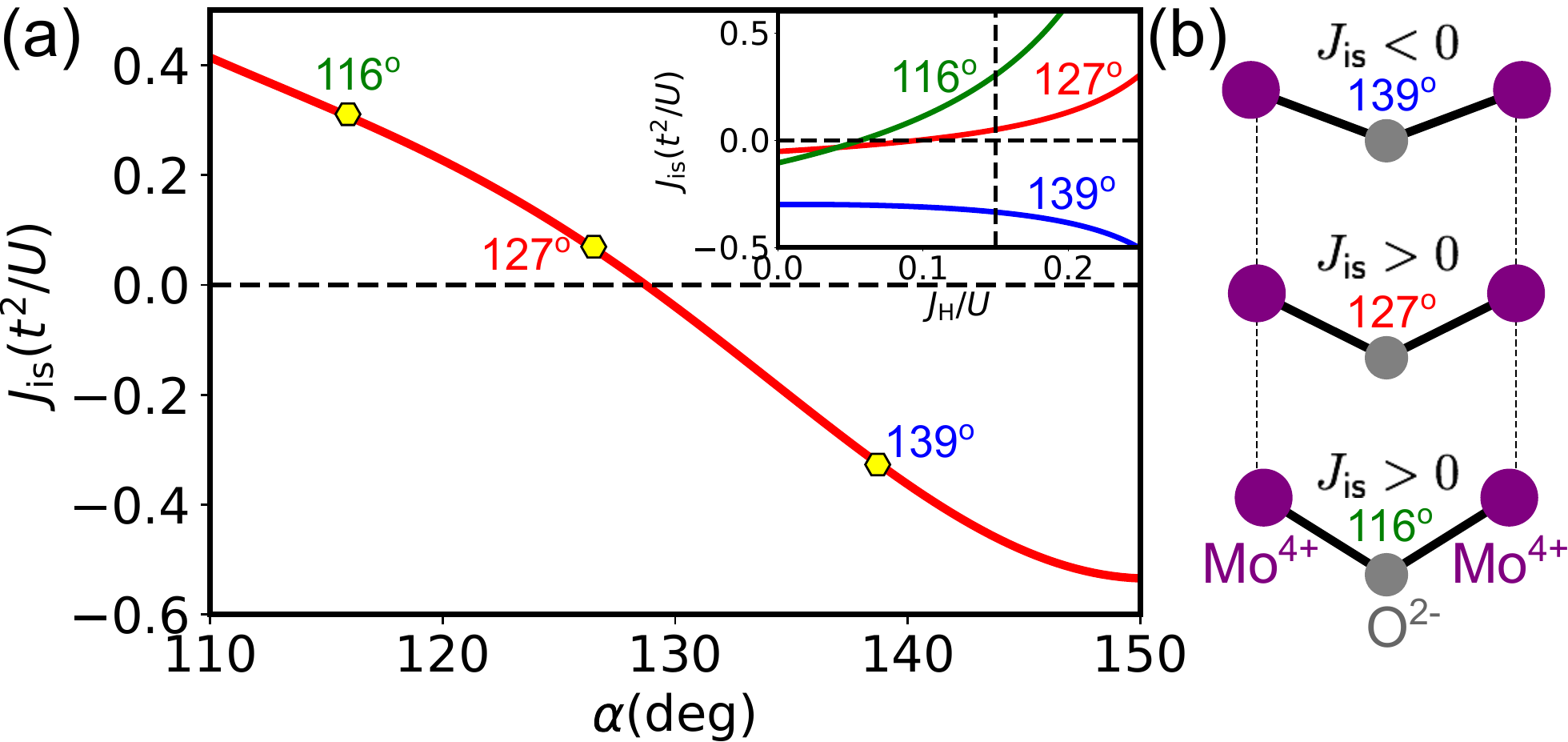}
\caption{\footnotesize{
Dependence of the Ising exchange interaction, $J_{\sf is}$, on the Mo-O-Mo bond angle $\alpha$.
(a) $J_{\sf is}$ is plotted for $J_{\sf H}/U=0.15$ and measured in units of $t^2/U$, where $t=V_{\sf pd\pi}^2/\Delta_{\sf dp}$ is the hopping integral along a Mo-O-Mo $\pi$-type bond \cite{slater54,supmat}.
For $\sigma$ bonds we use $V_{\sf pd\sigma} = -2.2V_{\sf pd\pi}$ \cite{harrison}.
The experimentally-determined bond angles for Y$_2$Mo$_2$O$_7$ are marked \cite{thygesen17} and $J_{\sf is}$ passes through zero close to the average bond angle, $\alpha_{\sf av}\approx 127^\circ$.
(Inset) Dependence of $J_{\sf is}$ on $J_{\sf H}/U$ for $\alpha = 116^\circ$ (green), $\alpha = 127^\circ$ (red) and $\alpha = 139^\circ$ (blue).
(b) Experimentally determined bond angles for Y$_2$Mo$_2$O$_7$ \cite{thygesen17}.
}}
\label{fig:Jis_alph}
\end{figure}

Due to the competition between the $\pi$ and $\sigma$ channels, the value of $J_{\sf is}$ is extremely sensitive to the bond angle, as shown in Fig.~\ref{fig:Jis_alph}{\red a}.
%
%
Serendipitously, the bond angle at which the exchange interaction cancels is close to $\alpha =  \alpha_{\sf av} =127^\circ$, both in our simple analytical calculations and also in more involved band structure calculations \cite{shinaoka13}.
Deviations of the Mo-O-Mo bond angle from its average value thus lead to very large relative changes in $J_{\sf is}$.

In addition to the trigonal distortion, it is also necessary to take into account the 2-in-2-out distortions of the Mo ions, and concomitant movement of the O octahedra, which leaves the Mo-O bond length and local crystal field almost invariant, but significantly changes the Mo-Mo distances and the Mo-O-Mo bond angles \cite{thygesen17}.
The result is three classes of Mo-Mo bonds with significantly different exchange couplings (see Fig.~\ref{fig:dpdhopping}{\red c}).
Combining the pair-distribution-function analysis with the calculations yields on each tetrahedron one short bond with $\alpha < \alpha_{\sf av}$ and $J_{\sf is}>0$, four medium bonds with $\alpha \approx \alpha_{\sf av}$ and $J_{\sf is}>0$ and one long bond with $\alpha > \alpha_{\sf av}$ and $J_{\sf is}<0$.
This conclusion remains robust, even if the bond angle deviations are smaller than the $\sim 11^\circ$ reported in \cite{thygesen17} and shown in Fig.~\ref{fig:Jis_alph}{\red b}.
The magnitude of $J_{\sf is}$ is significantly larger on the short and long bonds than the medium bonds, resulting in a very strong magnetoelastic coupling effect.


{\it Spin-lattice model}:
Next we consider the minimal effective model that captures the essence of the microscopic analysis.
Mo lattice sites are allowed to displace in a 2-in-2-out pattern (Fig.~\ref{fig:dpdhopping}{\red c}), corresponding to the triplet normal mode of a tetrahedron \cite{yamashita00,tchernyshyov02b,dimatteo05}.
The resultant bond angles are $\alpha = \alpha_{\sf av} + \delta \alpha$ on the long bond, $\alpha = \alpha_{\sf av}$ on the medium bonds and $\alpha = \alpha_{\sf av} - \delta \alpha$ on the short bond, with $\delta \alpha \geq 0$ (see Fig.~\ref{fig:Jis_alph}{\red b}).
For simplicity, $\delta \alpha$ is allowed to vary globally but not locally. 
Spins are constrained to point into or out of tetrahedra and described by $\sigma = \pm1$.
The resultant spin-lattice Hamiltonian is,
\begin{align}
\mathcal{H}_{\sf sl} = \sum_{\langle ij\rangle} \left[ \left( J_{\sf is}(\alpha_{\sf av}) - g \delta \alpha_{ij}   \right)\sigma_i \sigma_j + K\delta\alpha_{ij}^2 \right],
\label{eq:Hsl}
\end{align}
where $J_{\sf is}(\alpha_{\sf av})$ is the superexchange interaction for undistorted bonds, $g= \partial J_{\sf is}/\partial \alpha |_{\alpha=\alpha_{\sf av}}$ describes at first order the angular dependence of the interaction, $\delta \alpha_{ij} \in \{-\delta \alpha, 0, \delta \alpha\}$ and $K$ is the elastic energy cost of distorting the lattice.

\begin{figure}[t]
\centering
\includegraphics[width=0.4\textwidth]{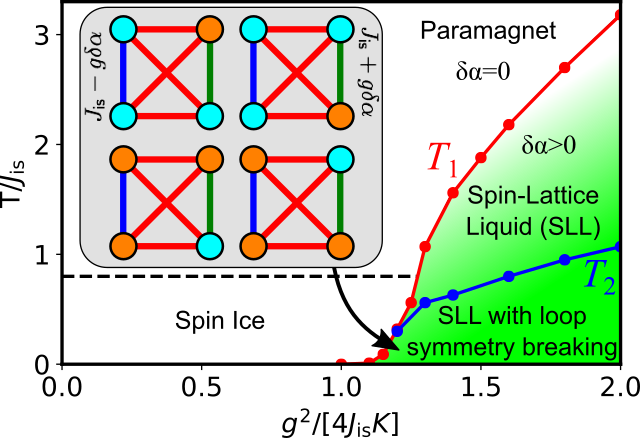}
\caption{\footnotesize{
Phase diagram of $\mathcal{H}_{\sf sl}$ [Eq.~\ref{eq:Hsl}] from Monte Carlo simulation.
For $g^2/(4K)<J_{\sf is}$ there is no lattice distortion and a crossover (marked by a dashed line) to spin ice behaviour occurs at low $T$.
For $g^2/(4K)>J_{\sf is}$ an incoherent lattice distortion occurs ($\delta \alpha >0$) and a spin-lattice liquid (SLL) forms below a first-order transition (red).
At lower $T$ a likely second-order transition (blue) picks out loops of length $l_{\sf loop}=4m+4$, $m=1,2,\cdots$ \cite{supmat}.
(Inset) Four degenerate 3-1 ground states of a tetrahedron in the SLL state with a fixed lattice distortion (short bond in green and long bond in blue), and with spins pointing in (orange) and out (blue).
}}
\label{fig:MEphasediag}
\end{figure}

The nature of the ground state can be understood starting from a single tetrahedron, and depends on the ratio $g^2/(4KJ_{\sf is}(\alpha_{\sf av}))$.
For $g^2/(4K)<J_{\sf is}(\alpha_{\sf av})$ it is energetically favourable to have $\delta \alpha(T=0)=\delta \alpha_{\sf gs} = 0$ combined with a 2-in-2-out spin configuration on each tetrahedra, and as a result a spin ice forms at low $T$.
For $g^2/(4K)>J_{\sf is}(\alpha_{\sf av})$ the lowest energy is reached with $\delta \alpha_{\sf gs} = g/(2K)$, and writing $J_{\sf is}(\alpha) = J_{\sf is}(\alpha_{\sf av}) \pm \delta J_{\sf is}$, one finds $\delta J_{\sf is} > 2J_{\sf is}(\alpha_{\sf av})$ on long and short bonds.
For a given lattice distortion a tetrahedron has four degenerate low-energy 3-1 spin configurations, where 3 spins point in and 1 out or vice versa, as shown in the inset of Fig.~\ref{fig:MEphasediag}
\footnote{The mechanism by which a 3-1 ground state appears is quite different from previous work, where a partially ordered monopole crystal has been proposed \cite{brooksbartlett14,jaubert15a,jaubert15b} and multiferroic behaviour discussed \cite{khomskii12}.}.
The combined manifold of spin and lattice configurations is extensive \cite{supmat}, and so a spin-lattice liquid may be realised at low $T$.

The full phase diagram can be explored using Monte Carlo simulation, and is shown in Fig.~\ref{fig:MEphasediag} (see supplemental material \cite{supmat} for details of the simulations).
We set $g/(2K)=0.2$, since this corresponds to the experimental findings at low $T$ ($\delta \alpha_{\sf gs}=0.2$ $\approx 11^\circ$) \cite{thygesen17}.
As expected, for $g^2/(4K)<J_{\sf is}(\alpha_{\sf av})$ there is no lattice distortion at any temperature, and, below a specific heat peak at $T\approx 0.8 J_{\sf is}$, the system shows the characteristic behaviour of spin ice
\footnote{Since analytical calculations show $J_{\sf is}(\alpha_{\sf av})$ is close to 0, we have checked that the phase diagram for $J_{\sf is}(\alpha_{\sf av})<0$ is qualitatively similar to that for $J_{\sf is}(\alpha_{\sf av})>0$. The main difference is that the value of $g^2/(4K)$ at which all-in-all-out order turns into a spin-lattice liquid is larger than the value at which a spin-ice becomes a spin-lattice liquid.}.

\begin{figure}[t]
\centering
\includegraphics[width=0.49\textwidth]{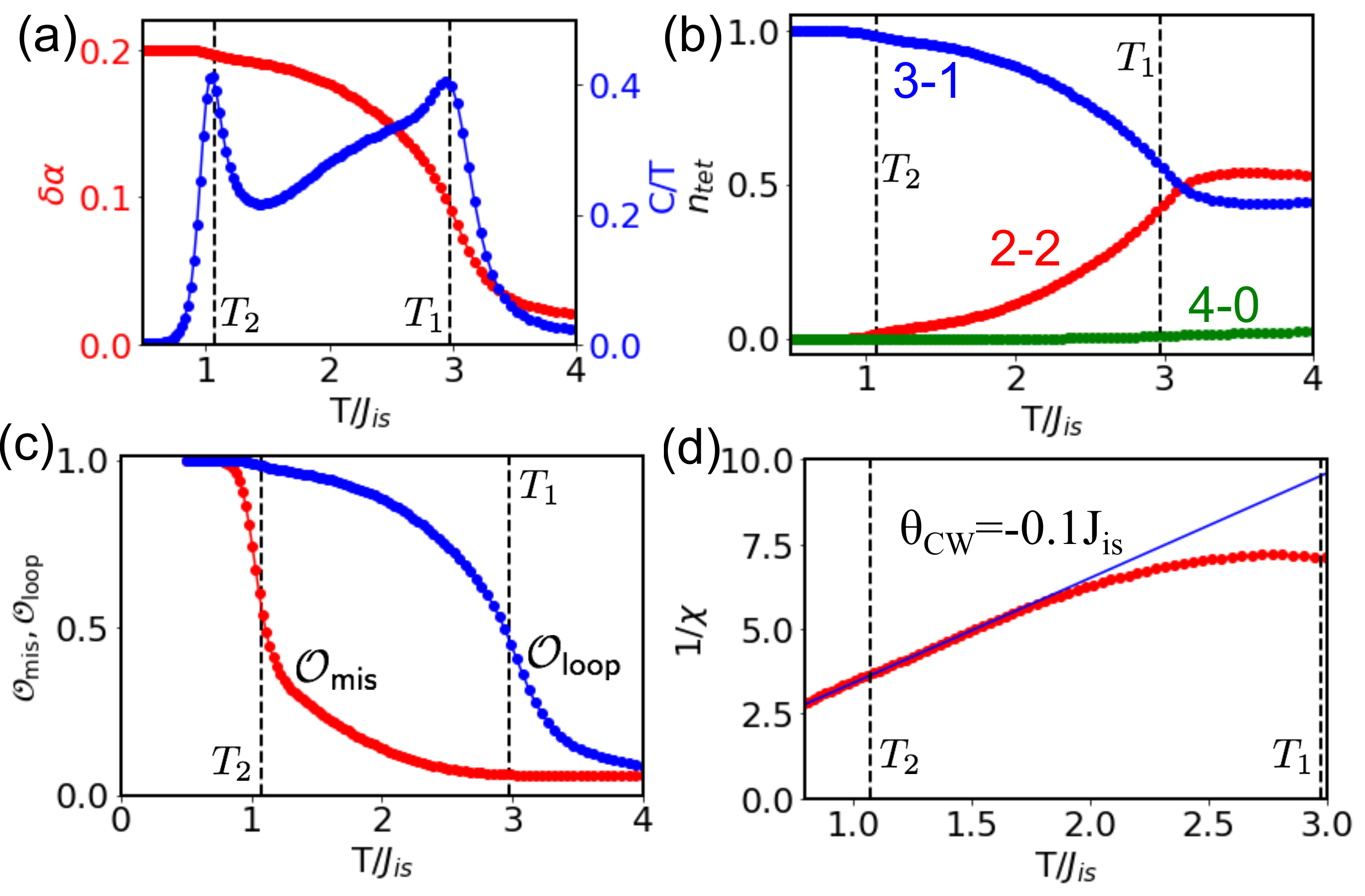}
\caption{\footnotesize{
Physical characteristics of $\mathcal{H}_{\sf sl}$ [Eq.~\ref{eq:Hsl}] from Monte Carlo simulation for $g^2/(4K)=2J_{\sf is}$.
Error bars are smaller than the point sizes.
(a) Lattice distortion, $\delta \alpha$, and heat capacity, $C/T$, showing two transitions.
Results are shown for $L=3$ (total number of spins $N=16L^3$), for which equilibration is possible across the lower-$T$ transition, and are consistent with simulations with larger $L$ \cite{supmat}.  
(b) Fraction of tetrahedra displaying 2-in-2-out (red), 3-1 (blue) or 4-0 (green) spin configurations. 
(c) Loop-based order parameters showing spin correlations on loops, $\mathcal{O_{\sf loop}}$, and mismatch between loops of length $l_{\sf loop}=4m+2$ and $l_{\sf loop}=4m+4$,  $\mathcal{O_{\sf mis}}$ \cite{supmat}.
(d) Inverse magnetic susceptibility.
The fit to $1/\chi \propto T-\theta_{CW}$ gives $\theta_{CW} = -0.1 J_{\sf is}$.
}}
\label{fig:da_CT_ntet}
\end{figure}

For $g^2/(4K)>J_{\sf is}(\alpha_{\sf av})$ there are two transitions as temperature is reduced, and this can be seen in simulations of the heat capacity shown in Fig.~\ref{fig:da_CT_ntet}{\red a}.
At $T=T_1$ there is a first-order liquid-gas-like transition in which $\delta \alpha(T)$ jumps discontinuously from $\delta \alpha(T>T_1) \approx 0$ to $\delta \alpha(T= T_1^-) \approx J_{\sf is}(\alpha_{\sf av})/g$.
Below this there are three classes of bonds, long and short bonds with coupling $J_{\sf is}(\alpha_{\sf av}) \pm g \delta \alpha(T)$ and medium bonds with coupling $J_{\sf is}(\alpha_{\sf av})$ (see inset to Fig.~\ref{fig:MEphasediag}).
For a fixed lattice configuration, the long and short bonds form non-intersecting loops that visit every lattice site and on which the sign of the coupling alternates.
Strong spin correlations form on these loops at the transition temperature, favouring alternating pairs of $\sigma=\pm 1$ \cite{supmat}.
The correlations increase on further decreasing $T$, and this is related to the increasing density of tetrahedra with a 3-1 spin configuration (see Fig.~\ref{fig:da_CT_ntet}{\red b}).
The loops themselves are not static, but constantly rearranging themselves subject to the 2-in-2-out constraint on the lattice displacements.
The distribution of loop lengths follows the same power-law scaling as is found in spin-ice \cite{jaubert11}, despite the interaction of lattice displacements via spin degrees of freedom (see supplemental material \cite{supmat} for more details about loops).

At lower $T$ there is a transition at $T=T_2$ that is likely second order, and in which the system excludes loops with length $l_{\sf loop}=4m+2$ ($m=1,2,...$) in favour of those with length $l_{\sf loop}=4m+4$, which we call loop symmetry breaking (see Fig.~\ref{fig:da_CT_ntet}{\red c} and \cite{supmat}).
This occurs because loops with length $l_{\sf loop}=4m+2$ cannot simultaneously minimise the energy of every bond,  as they are constrained to have an odd number of unsatisfied bonds, while loops with length $l_{\sf loop}=4m+4$ can satisfy all the bonds (or have an even number of unsatisfied bonds) \cite{supmat}.
Below $T=T_2$ essentially all tetrahedra adopt a 3-1 spin configuration (Fig.~\ref{fig:da_CT_ntet}{\red b}) and the energy saturates, but the number of allowed configurations remains extensive.

The essence of the resultant spin-lattice liquid state is that it is built by nonperturbatively coupling two classical spin liquids, which in the present case are both spin ice.
As with classical spin liquids, spin-lattice liquids can be described by a coarse-grained gauge theory, and, while this inherits some of the structure of the underlying spin-liquid gauge theories \cite{henley10,castelnovo12}, the nonperturbative coupling provides additional constraints on the allowed configurations.
A detailed exploration of such an effective field theory is left for future work.

%


{\it Relation to experiment}:
In order to determine an approximate value of the control parameter $g^2/(4KJ_{\sf is}(\alpha_{\sf av}))$ for Y$_2$Mo$_2$O$_7$, the calculated value of $g/J_{\sf is}(\alpha_{\sf av}) \sim 20$ can be combined with the experimentally determined $\delta\alpha_{\sf gs}= g/(2K) \approx 0.2$ \cite{thygesen17}, to give $g^2/(4KJ_{\sf is}(\alpha_{\sf av})) \sim 2$.
This places the material on the right of the phase diagram in Fig.~\ref{fig:MEphasediag}, firmly within the spin-lattice liquid region. 

The predictions of the model are predominantly aimed at understanding the intermediate temperature regime of Y$_2$Mo$_2$O$_7$, above the spin-glass freezing temperature, $T_{\sf f}$, but below the energy of the first crystal-field excitation, $\Delta E$.
This should be compared with the model in the region $T_2<T<T_1$.
Here, the discreteness of the Ising degrees of freedom, coupled with the discrete distortions of the Mo ions provide an explanation for the multiple peaks observed in NMR spectra of Y nuclei \cite{keren01}.
It is also consistent with the finding in $\mu$sr and NMR $1/T_2$ measurements that the lattice distortions are temperature dependent and respond strongly to magnetic field \cite{sagi05,ofer10}.

At lower temperatures the degeneracy of the model means that other subdominant effects are likely to play a role.
To understand the experimentally observed spin-glass region ($T \leq T_{\sf f}$) it is likely necessary to take into account some combination of disorder, further-neighbour interactions and additional exchange couplings arising from higher-energy crystal-field levels.
In particular, small local changes in the Mo-O-Mo bond angle, (i.e. in $\delta \alpha$) will result in additional disorder in the exchange couplings. 
One promising feature of the model is that its highly-correlated low-temperature dynamics likely make it very susceptible to spin-glass freezing, particularly in the vicinity of $T_2$, where Monte-Carlo simulations become difficult to equilibrate
\footnote{Using Monte Carlo to simulate the effect of disorder, and test whether it drives a spin-glass transition at temperatures close to $T_2$, is challenging, since it is very difficult to equilibrate the simulations due to the non-local nature of the dynamics and the strong correlation between spin and lattice degrees of freedom. We leave such a detailed investigation to future work.}.

At high temperatures ($T\sim \Delta E \sim 300$K) the excited crystal-field levels become thermally populated, and therefore the approximation of being in the low-energy doublet breaks down.
Since $\Delta E$ is smaller than the temperature at which NMR linewidth extrapolations determine that the lattice distortion dies away ($\sim 430$K) \cite{keren01}, the first-order transition predicted by the model is likely smeared out due to the additional local degrees of freedom and the opening of new hopping pathways.

A particular mystery in Y$_2$Mo$_2$O$_7$ is why susceptibility measurements show a negative (AFM) Curie-Weiss temperature of $\theta_{\sf cw} \approx -200$K in the high temperature range $500-800$K \cite{gardner04}, a different Curie-Weiss temperature of $\theta_{\sf cw} \approx -41$K in the intermediate range $50-300$K \cite{silverstein14} but ring features surrounding the $\Gamma$ point in low-temperature neutron-scattering measurements that can only be explained by including FM coupling \cite{silverstein14}.
Our model provides a resolution by proposing a mechanism by which AFM and FM interactions can coexist.
Simulations in the temperature range $T_2<T<T_1$ show an emergent Curie-Weiss behaviour with $\theta_{\sf cw} <0$ (see Fig.~\ref{fig:da_CT_ntet}{\red d}), despite the presence of strong correlations.
Above $T\sim 300K$ the local electronic state is no longer confined to the ground state doublet, and a change in the Curie-Weiss behaviour is unsurprising. 

One way to test for the existence of a spin-lattice liquid state in Y$_2$Mo$_2$O$_7$ would be to use magnetic field to control the lattice or uniaxial pressure to control the magnetism.
Large magnetic fields will suppress the 2-in-2-out lattice displacements by selecting spin configurations (either 2-in-2-out or 3-1) that are incompatible with the displacement.
Intermediate fields will have a more subtle effect, splitting the low-energy spin-lattice configurations, and could be an interesting direction for further study.


In conclusion, we have proposed the idea of a spin-lattice liquid state, arising from strong coupling between two ``spin'' liquids, one associated with the spin-orbital and the other with the lattice degrees of freedom.
Furthermore, we have shown that it may be realised in Y$_2$Mo$_2$O$_7$ and related compounds, driven by a giant magnetoelastic coupling effect.
According to our theory, the strength of the magnetoelastic coupling is the main thing that distinguishes the molybdate pyrochlores from $f$-electron spin-ice compounds such as Dy$_2$Ti$_2$O$_7$ \cite{bramwell01}.
Finally, we end with the hope that other materials may be found, in which the dance of spin and lattice extends down to even lower temperature.


{\it Acknowledgments.}   
We thank Reinhard Kremer, Ludovic Jaubert, Joe Paddison and Hidenori Takagi for useful comments on the manuscript.
%


\bibliographystyle{apsrev4-1}
\bibliography{bibfile}


\onecolumngrid
\clearpage
\begin{center}
\textbf{\large Supplemental Materials: Giant magnetoelastic-coupling driven spin-lattice liquid state in molybdate pyrochlores }
\end{center}
\setcounter{equation}{0}
\setcounter{figure}{0}
\setcounter{table}{0}
\setcounter{page}{1}
\makeatletter
\renewcommand{\theequation}{S\arabic{equation}}
\renewcommand{\thefigure}{S\arabic{figure}}


\section{Local electronic state of $d^2$ ions}
\label{sec:localelec}


Here we consider in more detail the local electronic state of an ion with two electrons in the $d$-shell ($d^2$ configuration), residing in a trigonally distorted oxygen octahedra and subject to spin-orbit coupling. 
The largest energy scale is the cubic crystal-field splitting of the $4d$ electron states, and as a result we consider only the $t_{\sf 2g}$ levels in which the $d$ electron orbitals point in between the surrounding oxygen sites \cite{fazekas}.
Further splitting occurs due to Hund's coupling, which favours parallel spin alignment, resulting in a low-energy manifold of states labelled by spin $S=1$ and effective angular momentum $L_{\sf eff}=1$.
The 9-fold degeneracy of these states is split by the combination of a trigonal flattening of the surrounding oxygen octahedra and spin orbit coupling, and we consider the local Hamiltonian,
\begin{align}
\mathcal{H}_{\sf loc} = 
\Delta_{\sf trig} \left[ \frac{2}{3} - (L^{\sf z^\prime}_{\sf eff})^2 \right] 
-\lambda_{\sf SO} {\bf S}\cdot {\bf L}_{\sf eff},
\label{eq:Hloc}
\end{align}
where ${\bf S}$ and ${\bf L}_{\sf eff}$ are measured relative to the axis of the trigonal distortion, which points into/out-of Mo tetrahedra and defines the local $z^\prime$ axis (see Fig.~1 in the main text).

\begin{figure}[h]
\centering
\includegraphics[width=0.6\textwidth]{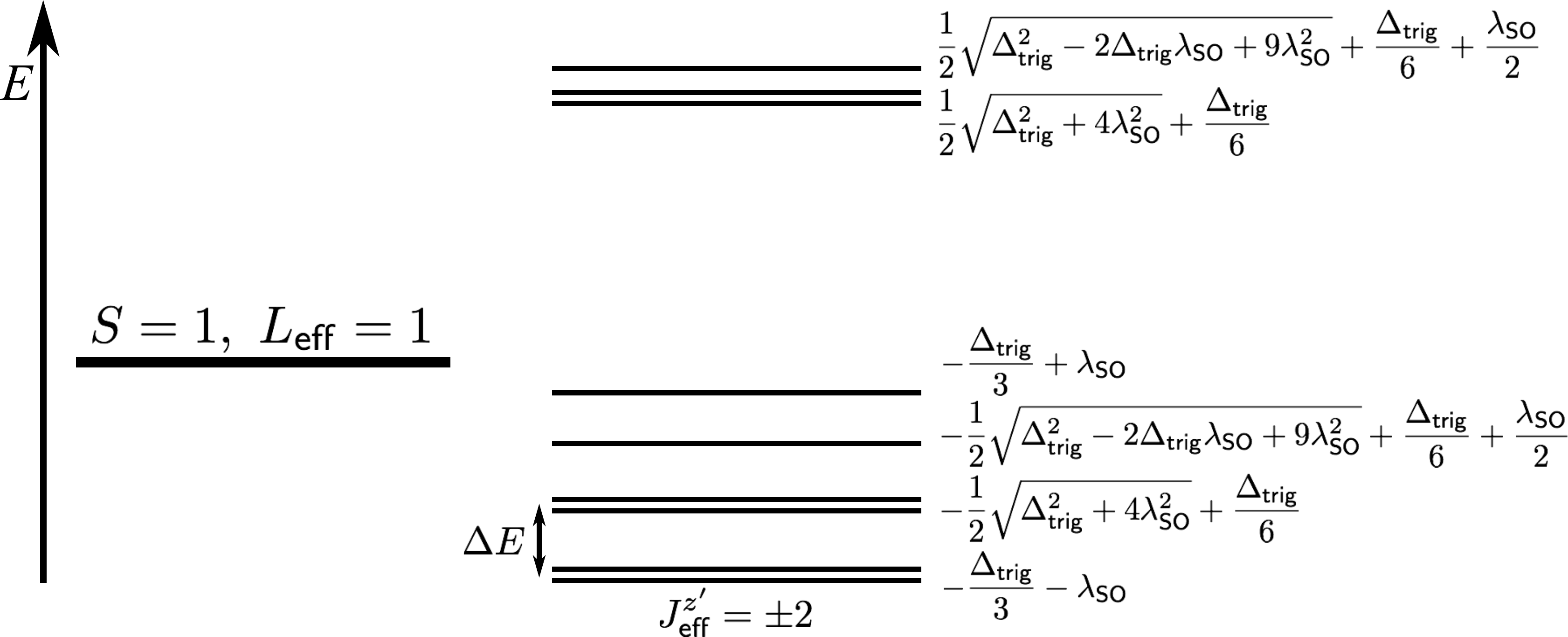}
\caption{\footnotesize{
Splitting of the 9-fold degenerate $S=1$, $L_{\sf eff}=1$ states by trigonal flattening of the oxygen octahedra and spin orbit coupling, according to $\mathcal{H}_{\sf loc}$ [Eq.~\ref{eq:Hloc}].
Levels are plotted for $\lambda_{\sf SO} = \Delta_{\sf trig}/4$ and energies are measured relative to the initial unsplit level.
}}
\label{fig:eleclevels}
\end{figure}

Diagonalisation of $\mathcal{H}_{\sf loc}$ gives the energy levels shown in Fig.~\ref{fig:eleclevels}.
In general, the presence of a trigonal distortion means that the levels cannot be labelled by their total angular momentum, ${\bf J}_{\sf eff} = {\bf S}+{\bf L}_{\sf eff}$.
However, the lowest energy doublet is independently an eigenstate of both parts of $\mathcal{H}_{\sf loc}$ and can therefore be written in terms of the total angular momentum as $|J_{\sf eff}=2, J_{\sf eff}^{z^\prime}=\pm2 \rangle = |S^{z^\prime}=\pm1,L_{\sf eff}^{z^\prime} =\pm1 \rangle$.

In the main text we consider this lowest-energy doublet to be the only accessible state at low temperature, and this is valid as long as $T \ll \Delta E =  \lambda_{\sf SO} [1-\lambda_{\sf SO}/\Delta_{\sf trig} +\dots]$, where $k_B=1$ and the energy difference between the levels has been expanded in terms of $\lambda_{\sf SO}/\Delta_{\sf trig}$.
In Y$_2$Mo$_2$O$_7$ band structure calculations have estimated that $\Delta_{\sf trig}\approx160$meV and $\lambda_{\sf SO} \approx 40$meV \cite{shinaoka13}, in reasonable agreement with the $\lambda_{\sf SO} \approx 58$meV found experimentally for isolated Mo$^{4+}$ ions \cite{abragam}.
In consequence one finds $\Delta E \approx 30$meV$\approx 300$K.


\section{Superexchange Hamiltonian on the pyrochlore lattice}
\label{sec:superexchange}


Here we derive the superexchange Hamiltonian used in the main text.
First we consider the idealised case of a pyrochlore lattice surrounded by regular oxygen octahedra.
While this is never exactly the case in materials such as Y$_2$Mo$_2$O$_7$, due to a tension between the oxygen octahedra surrounding the Mo ions and the oxygen cubes surrounding the Y ions, it serves as a useful starting point.
We then go on to consider the effect of trigonally disorting the oxygen octahedra, and show that the distortion can have a significant effect on the superexchange Hamiltonian.


\subsection{Superexchange in idealised case with regular octahedra}

\begin{figure}[h]
\centering
\includegraphics[width=0.3\textwidth]{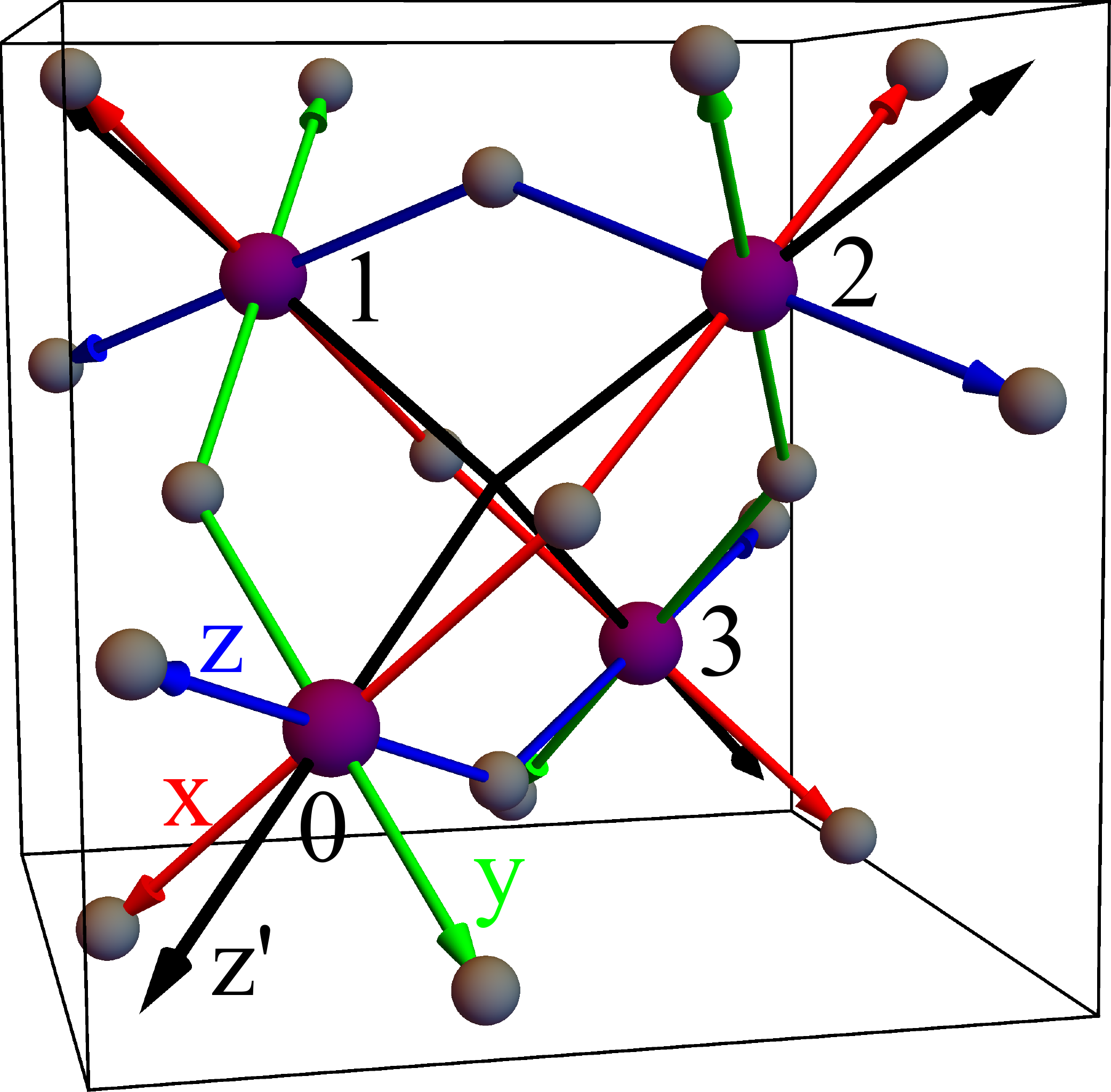}
\caption{\footnotesize{
Local cubic coordinate systems for a pyrochlore lattice surrounded by regular oxygen octahedra.
The 4 basis sites of the pyrochlore lattice (purple) form a tetrahedra and are surrounded by corner-sharing oxygen octahedra (grey).
Global (crystallographic) coordinates follow the axes of the bounding cube, while local coordinates $x$ (red), $y$ (green) and $z$ (axes) are site dependent.
Black lines show the local anisotropy $z^\prime$ axes associated with a trigonal octahedral distortion.
}}
\label{fig:tetrahedra}
\end{figure}

In the Y$_2$Mo$_2$O$_7$ structure, oxygen octahedra surrounding each of the 4 basis sites of the Mo pyrochlore lattice have different orientations.
As such it is useful to set-up a local cubic coordinate system for each site, as shown in Fig.~\ref{fig:tetrahedra}.
Rotation matrices from local cubic to global cubic (i.e. crytallographic) coordinates are given by,
\begin{align}
R^T_0 &= \left(
\begin{array}{ccc}
-2/3 & 1/3 & -2/3  \\
1/3 & -2/3 & -2/3 \\   
-2/3 & -2/3 & 1/3  \\
\end{array}
\right), \quad
R^T_1 = \left(
\begin{array}{ccc}
-2/3 & 1/3 & -2/3  \\
-1/3 & 2/3 & 2/3 \\   
2/3 & 2/3 & -1/3  \\
\end{array}
\right), \nonumber \\
R^T_2 &= \left(
\begin{array}{ccc}
2/3 & -1/3 & 2/3  \\
1/3 & -2/3 & -2/3 \\   
2/3 & 2/3 & -1/3  \\
\end{array}
\right), \quad
R^T_3 = \left(
\begin{array}{ccc}
2/3 & -1/3 & 2/3  \\
-1/3 & 2/3 & 2/3 \\   
-2/3 & -2/3 & 1/3  \\
\end{array}
\right).
\end{align}

We consider oxygen mediated $d$-$p$-$d$ hopping between neighbouring sites.
The $d$-$p$ hopping matrices taking $(d_{yz}, d_{zx},d_{xy})$ to $(p_x,p_y,p_z)$ are given by \cite{slater54,harrison} (see also Ref.~\cite{pesin10}),
\begin{align}
P_{\sf x}^\pm = \pm V_{pd\pi} 
\left(
\begin{array}{ccc}
0 & 0 & 0  \\
0 & 0 & 1 \\   
0 & 1 & 0  \\
\end{array}
\right), \quad
P_{\sf y}^\pm = \pm V_{pd\pi} 
\left(
\begin{array}{ccc}
0 & 0 & 1  \\
0 & 0 & 0 \\   
1 & 0 & 0  \\
\end{array}
\right), \quad
P_{\sf z}^\pm = \pm V_{pd\pi} 
\left(
\begin{array}{ccc}
0 & 1 & 0  \\
1 & 0 & 0 \\   
0 & 0 & 0  \\
\end{array}
\right),
\label{eq:Pmatrices}
\end{align}
where $V_{pd\pi}$ parametrises the $\pi$-bond hopping, the $\pm$ refers to a positive or negative oxygen coordinate and ${\sf x,y,z}$ labels the axis on which the oxygen lies.
On ``integrating out'' the oxygen orbitals one finds the $d$-$d$ hopping matrices,
\begin{align}
\Lambda_{ij} = 
t P_{ij}R_i R^T_j P_{ji} ,
\label{eq:hop_PRRP}
\end{align}
where $t= V_{pd\pi}^2/\Delta_{pd}$, $\Delta_{pd}$ gives the energy difference between the $d$ and $p$ orbitals and $P_{ij}$ is chosen appropriately for the bond in question.
For the regular octahedra shown in Fig.~\ref{fig:tetrahedra}, this gives the bond-dependent hopping matrices,
\begin{align}
\Lambda_{13} = \Lambda_{20} =
 t  \left(
\begin{array}{ccc}
0 & 0 & 0  \\
0 & -\frac{1}{9} & \frac{8}{9} \\   
0 & \frac{8}{9} & -\frac{1}{9}  \\
\end{array}
\right), \quad
\Lambda_{10} = \Lambda_{32} =
 t  \left(
\begin{array}{ccc}
-\frac{1}{9} & 0 & \frac{8}{9}  \\
0 & 0 & 0 \\   
\frac{8}{9} & 0 & -\frac{1}{9}  \\
\end{array}
\right), \quad
\Lambda_{03} = \Lambda_{12} =
 t  \left(
\begin{array}{ccc}
-\frac{1}{9} & \frac{8}{9} & 0  \\
\frac{8}{9} & -\frac{1}{9} & 0 \\   
0 & 0 & 0  \\
\end{array}
\right),
\label{eq:hop_regoct}
\end{align}
and it can be seen that the dominant hopping process is off-diagonal in the local coordinates.
The nearest-neighbour hopping Hamiltonian is given by,
\begin{align}
\mathcal{H}_\textsf{hop} = \sum_{\langle ij \rangle}\sum_{\alpha, \beta, s}
\alpha_{is}^\dagger \Lambda_{ij}^{\alpha\beta} \beta_{j s} + \mathrm{h.c.},
\end{align}
with $\alpha,\beta \in \{ d_{\sf yz},d_{\sf zx},d_{\sf xy}\}$ measured in the local cubic coordinates and $s \in {\uparrow,\downarrow}$.

It is also necessary to specify the on-site Coulomb interaction, and in the local octahedral coordinates this is given by \cite{castellani78},
\begin{align}
\mathcal{H}_{\sf site} =& (U-J_H) \left[
\sum_{\alpha,\beta} n_{\alpha\uparrow} n_{\beta\downarrow}
+\sum_{\alpha < \beta} (n_{\alpha\uparrow} n_{\beta\uparrow} + n_{\alpha\downarrow} n_{\beta\downarrow})
\right]
-2J_H \sum_{\alpha < \beta} \left( \textbf{S}_\alpha \cdot \textbf{S}_\beta +\frac{3}{4} n_\alpha n_\beta \right) \nonumber \\
&+  J_H \left[ \sum_{\alpha} n_{\alpha\uparrow} n_{\alpha\downarrow}
+\sum_{\alpha < \beta} \left(
\alpha^\dagger_\uparrow \alpha^\dagger_\downarrow \beta_\downarrow \beta_\uparrow
+\beta^\dagger_\uparrow \beta^\dagger_\downarrow \alpha_\downarrow \alpha_\uparrow 
\right)
\right],
\end{align}
where $U$ is the on-site Coulomb repulsion between electrons in the same orbital, $J_{\sf H}$ is Hund's coupling, $n_{\alpha s} = \alpha\dg_s \alpha\pdg_s$, $n_\alpha = n_{\alpha\uparrow} + n_{\alpha\downarrow}$ and 
$\textbf{S}_\alpha = (\frac{1}{2}[\alpha\dg_\uparrow \alpha\pdg_\downarrow + \alpha\dg_\downarrow \alpha\pdg_\uparrow], 
\frac{1}{2i}[\alpha\dg_\uparrow \alpha\pdg_\downarrow + \alpha\dg_\downarrow \alpha\pdg_\uparrow]),
\frac{1}{2}[\alpha\dg_\uparrow \alpha\pdg_\uparrow - \alpha\dg_\downarrow \alpha\pdg_\downarrow])$.

A spin-orbital Hamiltonian can be derived by the standard method of treating $t^2/U$ as a small parameter and considering virtual hopping processes of the type $d^2d^2 \rightarrow d^3 d^1\rightarrow d^2d^2$ \cite{khaliullin01,horsch03,khaliullin05}.
This results in,
\begin{align}
\mathcal{H}_\textsf{SO} = \frac{t^2}{U} \sum_\gamma \sum_{\langle ij \rangle \parallel \gamma}
\left[
\left( \textbf{S}_i \cdot \textbf{S}_j +1 \right) \hat{J}_{ij}^{(\gamma)} + \hat{K}_{ij}^{(\gamma)}
\right],
\label{eq:HSO}
\end{align}
where $\gamma$ refers to the three types of bonds  (01 \& 23 ; 02 \& 13 ; 03 \& 12) shown in Fig.~\ref{fig:tetrahedra}, $\textbf{S}_i$ is a spin-1 operator on site $i$ measured in the global coordinate system,
\begin{align}
\hat{J}_{ij}^{(\gamma)}(\eta) = (1+2\eta R) \hat{A}_{ij}^{(\gamma)} - \eta r \hat{B}_{ij}^{(\gamma)} - \eta R\hat{C}_{ij}^{(\gamma)} , \quad 
\hat{K}_{ij}^{(\gamma)}(\eta) =2\eta R \hat{A}_{ij}^{(\gamma)} +2\eta r \hat{B}_{ij}^{(\gamma)} - (1+\eta R)\hat{C}_{ij}^{(\gamma)},
\end{align}
$\eta = J_H/U$, $R=1/(1-3\eta)$, $r=1/(1+2\eta)$ and,
\begin{align}
\hat{A}_{ij}^{(\gamma)} &= \left[ -\frac{14}{9} \tau_i^z \tau_j^z 
+ \frac{64}{81} \left( \tau_i^+ \tau_j^+ + \tau_i^- \tau_j^- \right) 
+ \frac{1}{81} \left( \tau_i^+ \tau_j^- + \tau_i^- \tau_j^+ \right) 
+\frac{65}{162} n_i n_j \right.\nonumber \\
&\left. -\frac{8}{81} \left( n_i(\tau_j^+ + \tau_j^-) + (\tau_i^+ + \tau_i^-)n_j \right) \right]^{(\gamma)} \nonumber \\
\hat{B}_{ij}^{(\gamma)} &= \left[ -\frac{14}{9} \tau_i^z \tau_j^z 
+ \frac{1}{81} \left( \tau_i^+ \tau_j^+ + \tau_i^- \tau_j^- \right) 
+ \frac{64}{81} \left( \tau_i^+ \tau_j^- + \tau_i^- \tau_j^+ \right) 
+\frac{65}{162} n_i n_j \right.\nonumber \\
&\left. -\frac{8}{81} \left( n_i(\tau_j^+ + \tau_j^-) + (\tau_i^+ + \tau_i^-)n_j \right) \right]^{(\gamma)} \nonumber \\
\hat{C}_{ij}^{(\gamma)} &= \left[ \frac{65}{81} (n_i+n_j)  -\frac{16}{81} \left( \tau_i^+ + \tau_i^- + \tau_j^+  + \tau_j^- \right) \right]^{(\gamma)},
\end{align}
where the orbital operators are bond dependent pseudospins and given by $\tau^z = \frac{1}{2} [ d_{\sf yz}^\dagger d_{\sf yz} - d_{\sf zx}^\dagger d_{\sf zx} ]$, 
$\tau^+ = d_{\sf yz}^\dagger d_{\sf zx}$,
$\tau^- = d_{\sf zx}^\dagger d_{\sf yz}$ and 
$n = d_{\sf yz}^\dagger d_{\sf yz} + d_{\sf zx}^\dagger d_{\sf zx}$ on 03 and 12 bonds and by related expressions on other bonds.


\subsection{The effect of trigonal distortion and spin-orbit coupling}


The effect of the trigonal distortion on the spin-orbital Hamiltonian is twofold, since it is necessary to take into account a modified hopping matrix, as well the way in which it combines with spin-orbital coupling to split the otherwise 9-fold degenerate local energy levels.

When considering trigonally distorted oxygen octahedra, we continue to use a local cubic coordinate system that is aligned with the idealised undistorted octahedra (see Fig.~1 in the main text).
As a consequence, oxygen ions no longer sit on the axes of the local coordinates, and thus the $d$-$p$ hopping matrices become more complicated.
For example, using the tables given in Ref.~\cite{slater54,harrison}, the $P_{\sf z}^+$ matrix given in Eq.~\ref{eq:Pmatrices} should be replaced by,
\begin{align}
P_{\sf z}^+ =
\left(
\begin{array}{ccc}
\sqrt{3} lmn V_{pd\sigma} -2 lmn V_{pd\pi}   & 
\sqrt{3} l^2n V_{pd\sigma} +n(1-2l^2) V_{pd\pi} & 
\sqrt{3} l^2m V_{pd\sigma} +m(1-2l^2) V_{pd\pi}  \\
\sqrt{3} m^2n V_{pd\sigma} +n(1-2m^2) V_{pd\pi} & 
\sqrt{3} lmn V_{pd\sigma} -2 lmn V_{pd\pi} & 
\sqrt{3} m^2l V_{pd\sigma} +l(1-2m^2) V_{pd\pi} \\   
\sqrt{3} n^2m V_{pd\sigma} +m(1-2n^2) V_{pd\pi} & 
\sqrt{3} n^2l V_{pd\sigma} +l(1-2n^2) V_{pd\pi} & 
\sqrt{3} lmn V_{pd\sigma} -2 lmn V_{pd\pi}  \\
\end{array}
\right),
\end{align}
where $\textbf{r}_\textsf{O}=(l,m,n)$ is the position of the oxygen ion in the local cubic coordinate system.
Other hopping matrices follow in a similar manner.
For Mo$^{4+}$ ions it has been determied that $V_{pd\sigma}/V_{pd\pi} \approx -2.2$, and in fact this ratio is only weakly ion dependent \cite{harrison}.

For trigonal distortions of the type found to exist in Y$_2$Mo$_2$O$_7$, the position of the oxygen ion on a 03 bond can be written as,
\begin{align}
\textbf{r}_\textsf{O} = \left( 
\frac{1}{\sqrt{2}} \sin \left[\frac{\alpha_{\sf cub}-\alpha}{2} \right], 
\frac{1}{\sqrt{2}} \sin \left[\frac{\alpha_{\sf cub}-\alpha}{2} \right],
-\cos  \left[\frac{\alpha_{\sf cub}-\alpha}{2} \right]
\right),
\end{align}
where $\alpha_{\sf cub} = 2\arctan2\sqrt{2} \approx 141^\circ$ is the bond angle in the case of regular octahedra and 
$\alpha$ is the bond angle of the trigonally distorted structure, which for the average $Fd\bar{3}m$ structure of Y$_2$Mo$_2$O$_7$ is given by $\alpha \approx 127^\circ$.
Inserting these values into Eq.~\ref{eq:hop_PRRP} for the $d$-$d$ hopping matrix, one finds for the 03 bond and for $\alpha = 127^\circ$,
\begin{align}
\Lambda_{03}(127^\circ) = 
 t  \left(
\begin{array}{ccc}
-0.63 & 0.36 & 0.0025  \\
0.36 & -0.63 & 0.0025 \\   
0.0025 & 0.0025 & 0.0029  \\
\end{array}
\right).
\end{align}
Comparison to Eq.~\ref{eq:hop_regoct} for the case of regular oxygen octahedra shows that the balance between diagonal and off-diagonal hopping of $d_{\sf yz}$ and $d_{\sf zx}$ orbitals has changed significantly, while hopping to/from $d_{\sf xy}$ orbitals remains unimportant.
For regular octahedra the off-diagonal $d_{\sf yz} \leftrightarrow d_{\sf zx}$ hopping dominates, while for trigonally distorted octahedra the diagonal $d_{\sf yz} \leftrightarrow d_{\sf yz}$ and $d_{\sf zx} \leftrightarrow d_{\sf zx}$ hopping becomes more important.
The reason for this is the appearance of a $\sigma$-bond hopping channel that competes with the original $\pi$-bond hopping channel as soon as the oxygen octahedra are not regular (see Fig.~1 in the main text).
Due to the relative importance of the $\sigma$ channel ($V_{pd\sigma}^2 \approx 4.8 V_{pd\pi}^2$) even a relatively small change in the bond angle away from the case of regular octahedra can significantly alter the hopping matrix, and this is at the heart of the giant magnetoelastic effect proposed in the main text.

The combination of the trigonal distortion and the spin-orbit coupling split the 9-fold degenerate Hund's-rule coupled $t_{2g}$ levels ($S=1$, $L_{\sf eff}=1$), resulting in a ground state doublet labelled by $J_{\sf eff}^{z^\prime}=\pm2$, where the $z^\prime$ axis points into or out of the tetrahedra and defines a trigonal coordinate system (see Fig.~\ref{fig:tetrahedra}).
If the energy gap between the doublet and the higher energy levels is considerably larger than the exchange coupling strengh, $\sim t^2/U$, then the higher energy levels can be ignored.
We assume this is the case and therefore project a modified version of $\mathcal{H}_\textsf{SO}$ [Eq.~\ref{eq:HSO}] that takes the bond distortion into account into the manifold of local $J_{\sf eff}^{z^\prime}=\pm2$ states.
Since none of the operators appearing in $\mathcal{H}_\textsf{SO}$ [Eq.~\ref{eq:HSO}] connect $J_{\sf eff}^{z^\prime}=2$ and $J_{\sf eff}^{z^\prime}=-2$, the Hamiltonian takes the simple Ising form,
\begin{align}
\mathcal{H}_{\sf is} = J_{\sf is}(\alpha,\eta) \sum_{\langle ij\rangle} \sigma_i \sigma_j,
\end{align}
where $\sigma = \pm1$ represents moments pointing into or out of tetrahedra along the local trigonal $z^\prime$ axis.

The value of $J_{\sf is}(\alpha,\eta)$ can be determined by comparing the different configurational energies of a pair of neighbouring moments, resulting in,
 \begin{align}
J_{\sf is}(\alpha,\eta) =& -\frac{1+r(\eta)}{6} \frac{t_{a_{1g}}(\alpha)^2}{U}
+ \left( \frac{7R(\eta)}{9} +\frac{r(\eta)}{6} - \frac{11}{18} \right)\frac{t_d(\alpha)^2}{U}
- \left( \frac{5R(\eta)}{9} +\frac{r(\eta)}{3} - \frac{2}{9} \right)\frac{|t_{od}(\alpha)|^2}{U} \nonumber \\
&- \left( -\frac{2R(\eta)}{9} +\frac{r(\eta)}{3} + \frac{5}{9} \right)\frac{|t_{a_{1g}e}(\alpha)|^2}{U},
\label{eq:JIs-eta}
\end{align}
where $t_{a_{1g}}(\alpha)$, $t_d(\alpha)$, $t_{od}(\alpha)$ and $t_{a_{1g}e}(\alpha)$ are hopping parameters in the trigonal coordinate system, defined for 03 bonds according to,
\begin{align}
\Lambda^\textsf{trig}_{ij}(\alpha) =  
\left(
\begin{array}{ccc}
t_{a_{1g}} & t_{a_{1g}e} & t_{a_{1g}e}^*  \\
t_{a_{1g}e}^* & t_d  & t_{od} \\   
t_{a_{1g}e} & t_{od}^*  & t_d   \\
\end{array}
\right)
=\textbf{U} \cdot \Lambda_{ij} \cdot \textbf{U}^\dagger, \quad
\textbf{U} = \frac{1}{\sqrt{3}}
\left(
\begin{array}{ccc}
1 & 1 & 1  \\
1 & \omega  & \omega^* \\   
1 & \omega^*  & \omega   \\
\end{array}
\right), \quad
\omega=e^{-\frac{2i\pi}{3}},
\end{align}
where the trigonal basis is $(a_{1g},e_{g+}^\prime,e_{g-}^\prime)$.
For other bonds the $\textbf{U}$ must be modified, but this does not affect the value of $J_{\sf is}(\alpha,\eta)$.
Fig.~2 in the main text is plotted using the expression for $J_{\sf is}(\alpha,\eta)$ given in Eq.~\ref{eq:JIs-eta}.


\section{Physical properties of $\mathcal{H}_{\sf sl}$}


Here we describe in more detail the properties of the spin-lattice Hamiltonian $\mathcal{H}_{\sf sl}$ [Eq.~1 in the main text].
%


\subsection{Monte Carlo simulation details}


We used Monte Carlo simulations to determine the properties of $\mathcal{H}_{\sf sl}$.
Monte Carlo updates were performed consecutively for spin and lattice degrees of freedom.
For the spin degrees of freedom a single spin-flip algorithm was employed.
Updates of the lattice degrees of freedom (displacements into or out of a tetrahedra) were split into two steps.
First a worm algorithm was used to move between lattice configurations in which each tetrahedra has a 2-in-2-out pattern \cite{jaubert-thesis}.
The worm was formed taking into account the interaction of the lattice with the (fixed) spin degrees of freedom, and could therefore be made rejection free.
Second, a global change to the magnitude of the lattice distortion, $\delta \alpha$, was proposed, and accepted or rejected according to the Metropolis criteria.
Parallel tempering was employed to further improve equilibration.

\begin{figure}[h]
\centering
\includegraphics[width=0.4\textwidth]{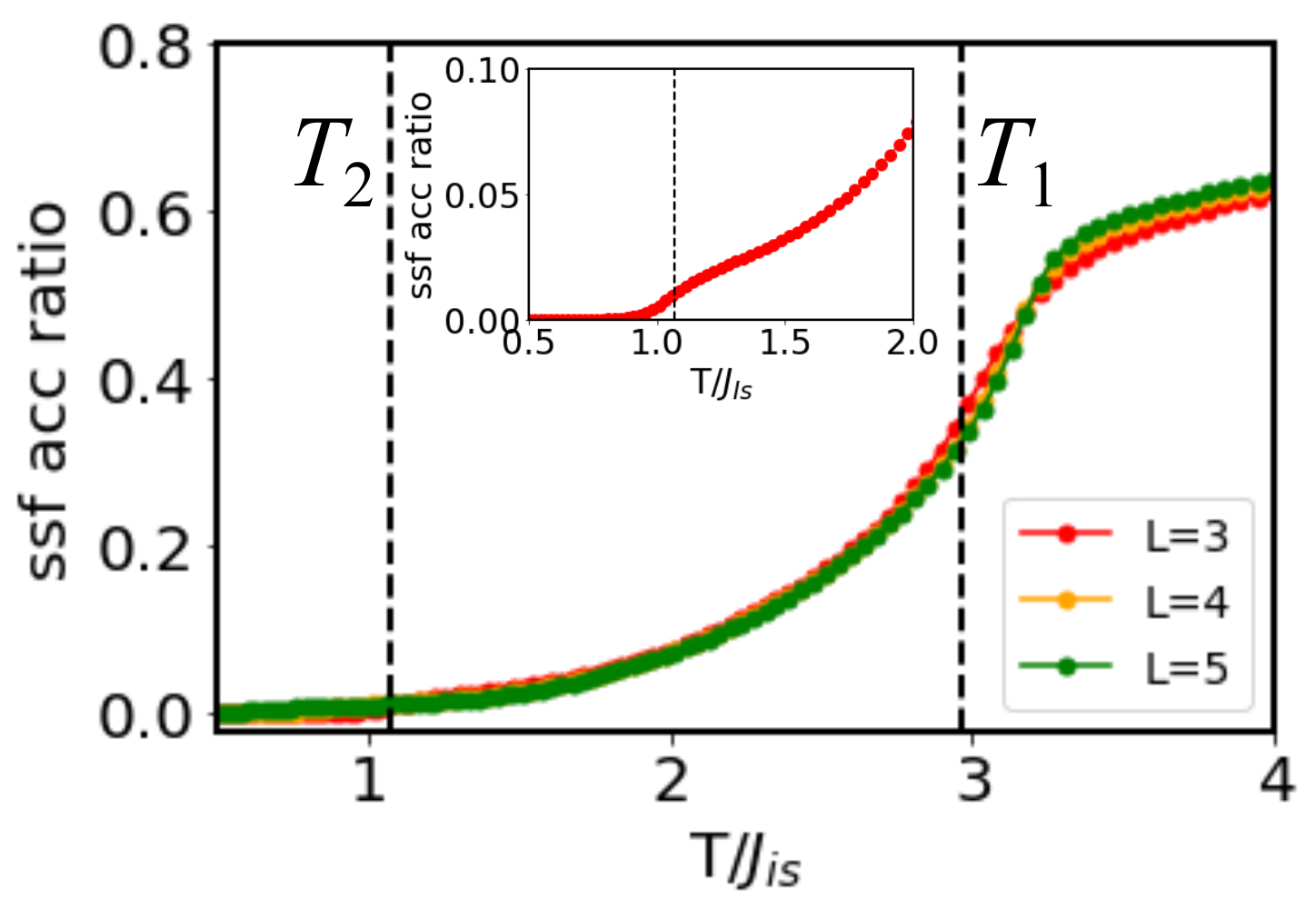}
\caption{\footnotesize{
Single spin-flip acceptance ratio for spin updates in Monte Carlo simulations of $\mathcal{H}_{\sf sl}$ [Eq.~1 in the main text], shown for $g^2/(4J_{\sf is}K)=2$.
The inset shows the vicinity of $T=T_2$ in more detail, where only small cluster sizes remain equilbrated across the transition.
}}
\label{fig:accrat}
\end{figure}

Typically a Monte Carlo step involved $N$ attempts to make a single spin flip, one worm step and 100 attempts to change $\delta \alpha$, with parallel tempering after each 10 Monte Carlo steps.
The number of spins was $N=16L^3$, where each unit cell has 16 sites, $L$ is the linear size of a cubic cluster and periodic boundary conditions were used.
Simulations were performed for $L=3,4,\dots ,12$ (i.e. $N=432, 1024, \dots ,27648$).
Equilibriation at the lower transition ($T=T_2$) proved difficult except for the smallest cluster sizes ($L=3,4$), due to the strong correlation between spin and lattice degrees of freedom. 
Overcoming this equilibration difficulty would almost certainly require a combined spin and lattice update.
Typical acceptance ratios of the single spin-flip algorithm are shown in Fig.~\ref{fig:accrat}, and it can be seen that they become very small in the neighbourhood of $T=T_2$.

%


\subsection{Thermodynamic properties}


The Monte Carlo simulations were used to determine the thermodynamic properties of $\mathcal{H}_{\sf sl}$ and thus map out the phase diagram shown in Fig.~3 of the main text.
\begin{figure}[h]
\centering
\includegraphics[width=0.7\textwidth]{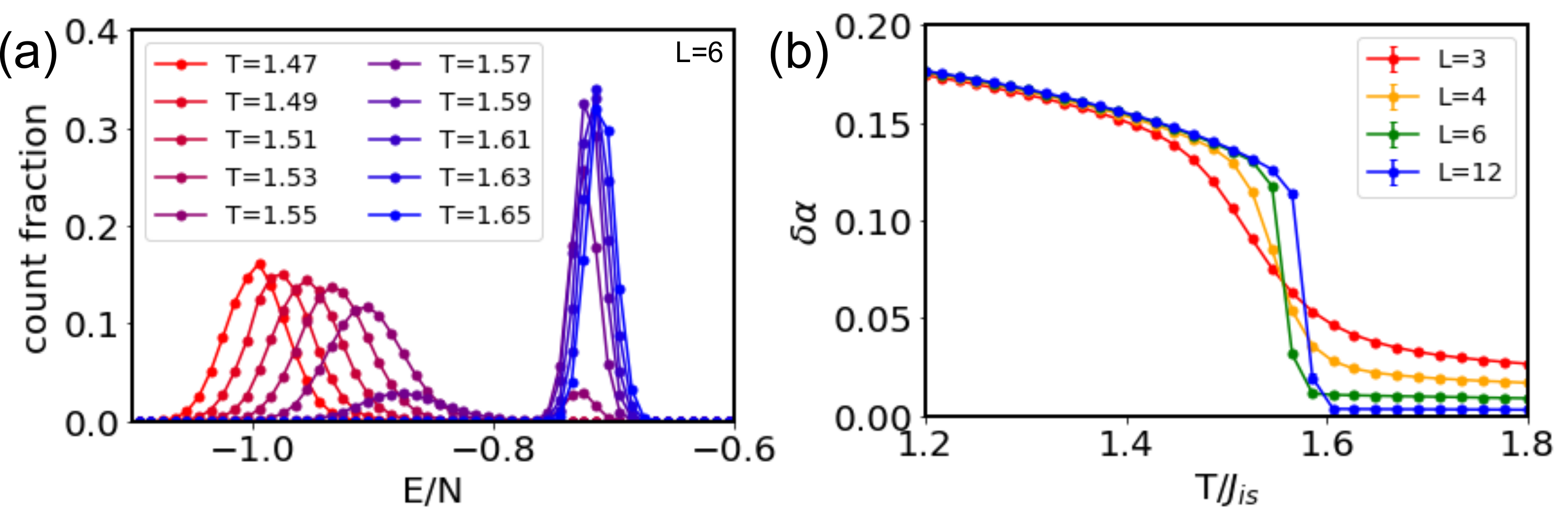}
\caption{\footnotesize{
Characteristics of a first-order liquid-gas type transition at $T=T_1$, shown for $g^2/(4J_{\sf is}K)=1.4$.
(a) Energy histogram analysis demonstrates a first-order transition, as can be seen from the double-peak structure.
The $L=6$ cluster has a transition at $T=1.57(2)J_{\sf is}$.
(b) The lattice displacement parameter, $\delta \alpha(T)$, for varying system sizes, showing a jump at $T=T_1$.
Error bars are smaller than the point size.
}}
\label{fig:Ehist}
\end{figure}

The first-order nature of the transition at $T=T_1$ can be ascertained from energy histogram analysis, as shown in Fig.~\ref{fig:Ehist}.
The double-peak structure is a sign of phase coexistence at the transition temperature.
That the transition is of the liquid-gas type can be seen from the behaviour of the lattice displacement parameter, $\delta \alpha$, which plays the role of density, and jumps discontinuously at the transition (see Fig.~\ref{fig:Ehist}).
It can also be seen in Fig.~\ref{fig:Ehist} that the transition temperature is relatively insensitive to system size.

The nature of the lower transition is more difficult to determine, since the Monte Carlo simulations encounter equilibration problems in the vicinity of the transition for larger system sizes.
Energy histogram analysis of small system sizes ($L=3,4,6$) appear to show a second-order transition.
However, finite size scaling analysis suffers from uncertainty due to the equilibration difficulties.
%


\subsection{Loops}


Here we explain in more detail the microscopic structure in the spin-lattice liquid phase (see Fig.~3 in the main text for the phase diagram).

\begin{figure}[h]
\centering
\includegraphics[width=0.4\textwidth]{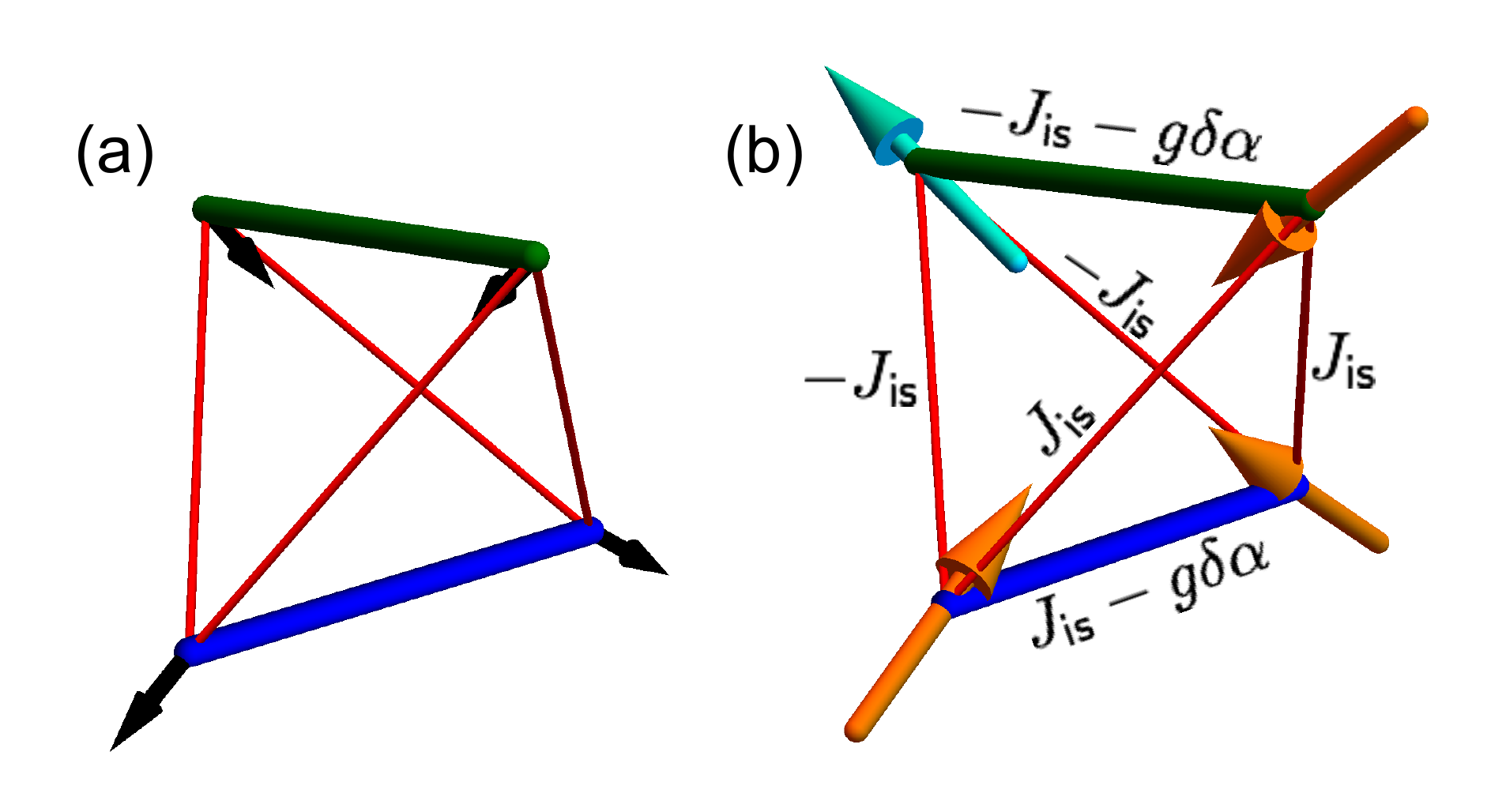}
\caption{\footnotesize{
A single tetrahedron showing lattice displacement and spin degrees of freedom.
(a) A 2-in-2-out lattice displacement of the ions (individual displacements shown by black arrows) results in one long bond (blue), four medium bonds (red) and one short bond (green).
Lattice displacements are fully specified by the bond colouring.
(b) Spins point into (orange) or out of (light blue) the tetrahedron, and for a given lattice displacement there are four lowest energy configurations with a 3-1 configuration, one of which is shown (see also Fig.~3 in the main text).
In the 3-1 spin configurations the exchange energies of the medium bonds cancel, and only the long and short bonds contribute to the total energy.
}}
\label{fig:tetrahedron}
\end{figure}

The building blocks of the pyrochlore lattice are tetrahedra, and in the spin-lattice liquid phase these have two ions displaced towards and two away from their centres.
Each tetrahedron thus has one long bond, four medium-length bonds and one short bond, and the long and short bonds are opposite one another (see Fig.~\ref{fig:tetrahedron}).
The effective spin model on a tetrahedron with a fixed lattice displacement consists, in the local Ising basis, of one ferromagnetic bond (long), four weakly antiferromagnetic bonds (medium) and one more strongly antiferromagnetic bond (short).
As a result, an isolated tetrahedron with fixed lattice displacement has a fourfold-degenerate spin ground state, with three spins pointing in and one out or vice versa (see Fig.~3 in the main text).
In these configurations, the four medium-length bonds host two satisfied interactions and two unsatisfied interactions, and therefore do not contribute to the energy.

\begin{figure}[h]
\centering
\includegraphics[width=0.9\textwidth]{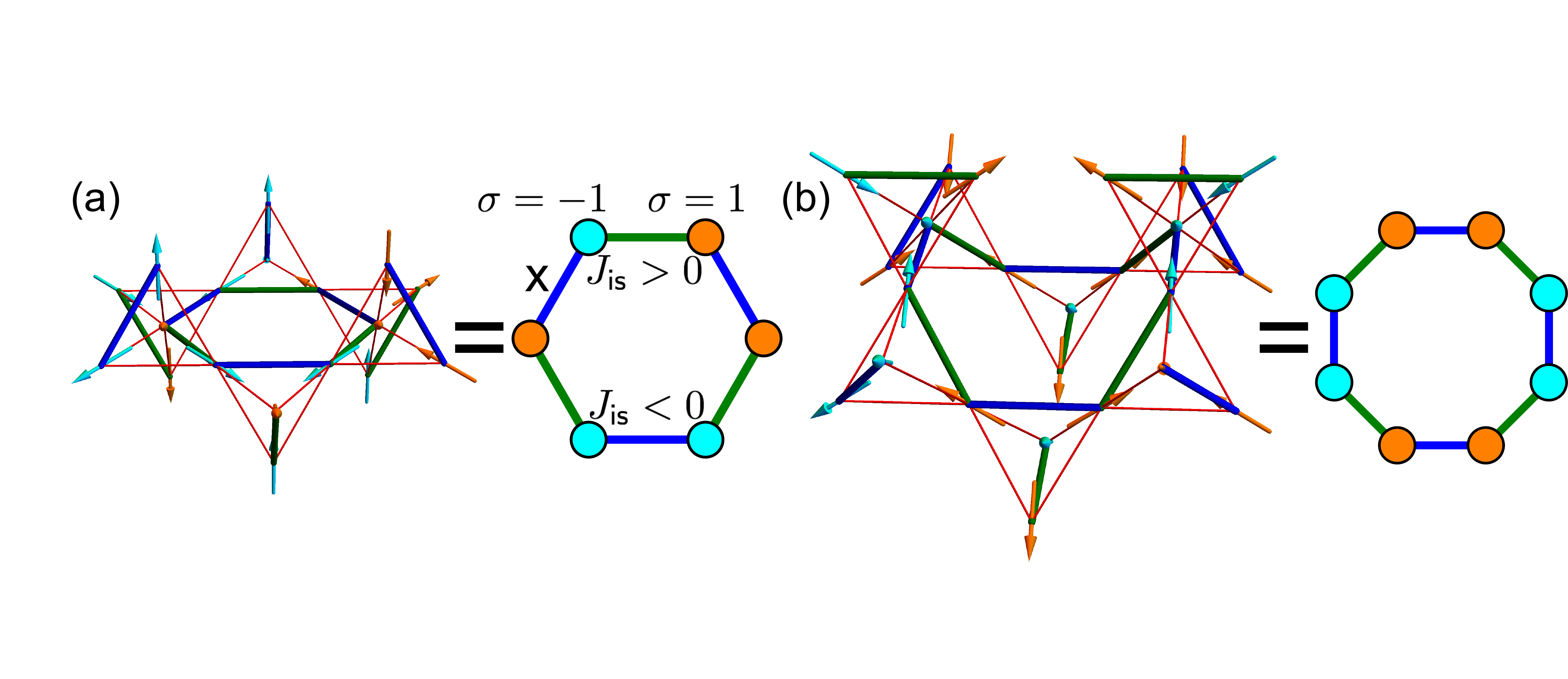}
\caption{\footnotesize{
Loops of alternating long (blue, $J_{\sf is}<0$) and short (green, $J_{\sf is}>0$) bonds, shown both on the pyrochlore lattice and as 2D projections.
(a) 6-site loops are frustrated in the sense that there must be at least one unsatisfied spin interaction on the loop (marked by a cross).
(b) 8-site loops can satisfy all the spin interactions.
}}
\label{fig:loops}
\end{figure}

On the pyrochlore lattice, the requirement that lattice displacements follow a 2-in-2-out rule results in the formation of  loops consisting of alternating long and short bonds.
Each lattice site is visited by exactly one loop, and the loops are non-intersecting and are closed in the presence of periodic boundary conditions.
The shortest possible loop has length $l_{\sf loop}=6$, and, due to the structure of the pyrochlore lattice, loops have to be even in length.
Some example loops are shown in Fig.~\ref{fig:loops}.

\begin{figure}[h]
\centering
\includegraphics[width=0.6\textwidth]{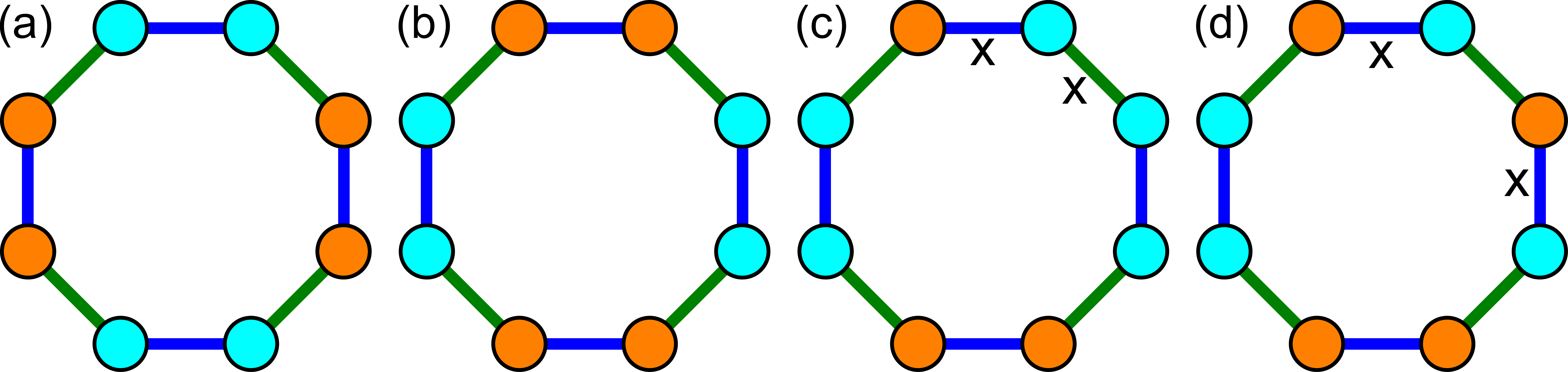}
\caption{\footnotesize{
Spin configurations on an 8-site loop.
(a) A ground-state spin configuration of the loop.
(b) Flipping all the spins on the loop gives the time-reversed ground-state configuration.
(c) Flipping one spin creates a pair of kinks (black crosses) and these are associated with tetrahedral configurations outside the 3-1 ground-state.
(d) Flipping a second spin moves the location of one of the kinks, and thus pairs of kinks can be separated.
}}
\label{fig:loopkinks}
\end{figure}

\begin{figure}[h]
\centering
\includegraphics[width=0.8\textwidth]{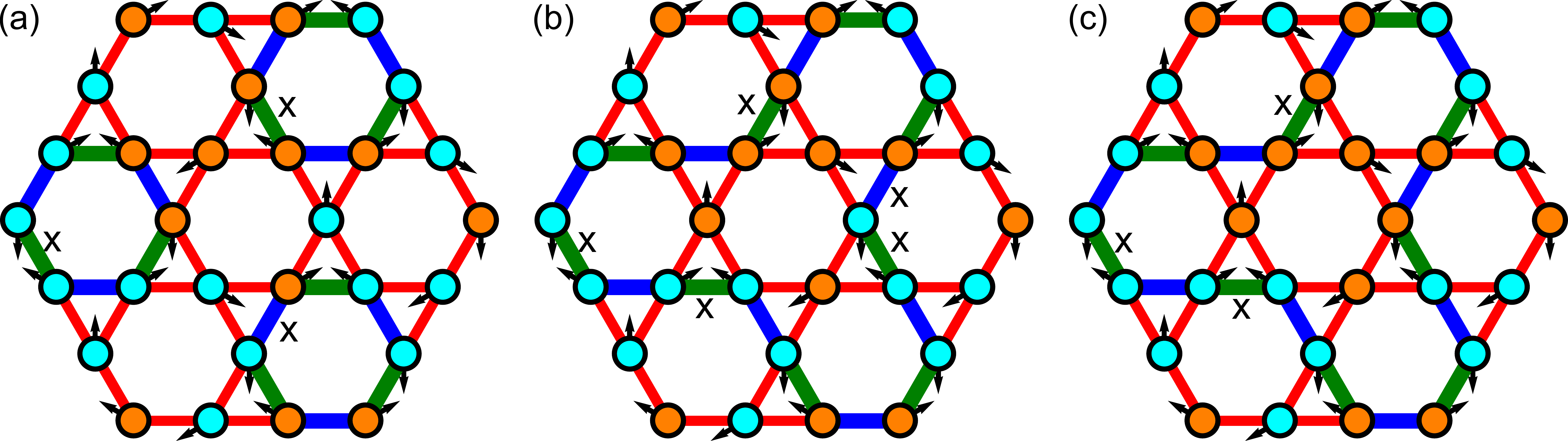}
\caption{\footnotesize{
An example of a local loop and spin update that leaves the energy invariant.
(a) Three 6-site loops in a kagom{\'e} plane of the pyrochlore lattice, with the planar projection of the lattice displacements shown as black arrows.
Each loop has one kink associated with an unsatisfied spin interaction (marked by black crosses).
(b) Reversing the lattice displacements on the central hexagon changes the loop structure from three 6-site loops to one 18-site loop.
Since there is no simultaneous change of the spins, the number of kinks is not conserved, and in this example the 18-site loop has five kinks, two more than the three 6-site loops. 
(c) Flipping a spin  removes two kinks and thus the energy (in the kagom{\'e} plane) of the new configuration is the same as that of the initial configuration.
}}
\label{fig:loopupdate}
\end{figure}

Configurations of the system are labelled by a combination of the loop structure (or equivalently the in/out lattice displacements) and the spin directions.
There is an extensive set of degenerate ground-state configurations in which all tetrahedra have a 3-1 (three in and one out or vice versa) spin configuration of the type shown in Fig.~\ref{fig:tetrahedron} (see also Fig.~3 in the main text).
These ground-state (gs) configurations consist of loops of length $l_{\sf loop} = 4+4m$ with $m=1,2,3,\dots$ (i.e. loops of length $l_{\sf loop} = 2+4m$ are excluded) and Ising spin configurations on the loops of $\boldsymbol{\sigma}^{\sf loop,gs} = (1,1,-1,-1,1,1,-1,-1,\dots)$.
There are two types of local move that leave the system within the ground state manifold. 
One possibility is that the spins on a loop are all flipped simultaneously, taking $\boldsymbol{\sigma}^{\sf loop,gs} \to -\boldsymbol{\sigma}^{\sf  loop,gs}$ (see Fig.~\ref{fig:loopkinks}).
A second possibility is that the bond displacements are exchanged around a closed set of medium-length bonds thus rearranging the loop structure of the lattice (see Fig.~\ref{fig:loopupdate} for an example of a non-ground-state loop update).
Typically it is simultaneously necessary to flip some spins to ensure that the reconstructed loops retain the correct $\boldsymbol{\sigma}^{\sf  loop,gs}$ structure.

One example of an excitation out of the ground state is the introduction of kinks into the otherwise periodic Ising configuration on a loop (see Fig.~\ref{fig:loopkinks}).
For loops of length $l_{\sf loop} = 4+4m$ the number of kinks has to be even, and each kink is associated with a 2-in-2-out (or potentially all-in-all-out) spin configuration on the relevant tetrahedron.
Excitations can also involve the creation of loops with length $l_{\sf loop} = 2+4m$, and the reason that these cost energy is that they have to contain at least one kink (see Fig.~\ref{fig:loops}{\red a}).
Further kinks can be created in pairs, and so the number of kinks is constrained to be odd.

\begin{figure}[h]
\centering
\includegraphics[width=0.35\textwidth]{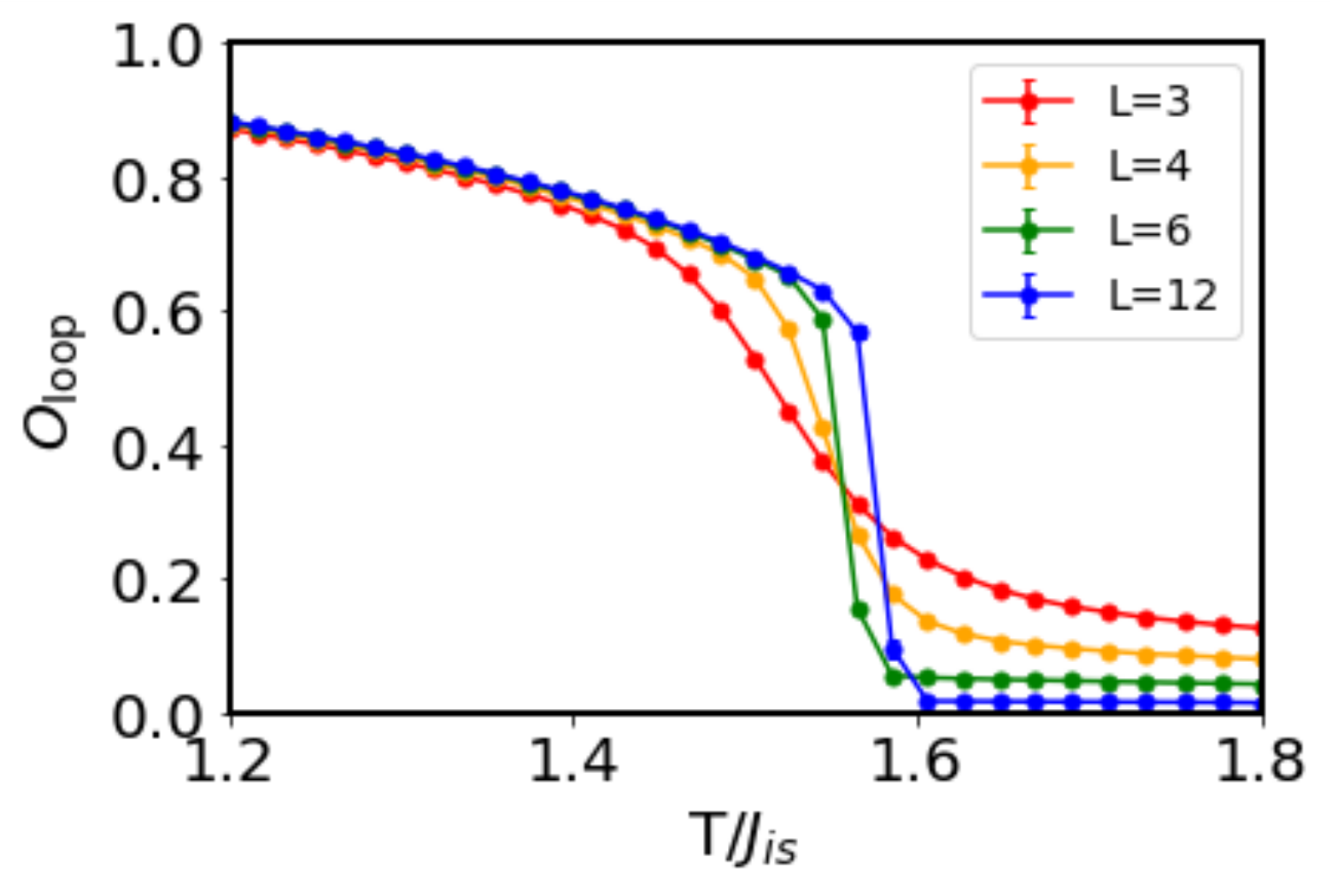}
\caption{\footnotesize{
The loop-spin order parameter, $\mathcal{O}_{\sf loop}$ [Eq.~\ref{eq:Oloop}], as measured by Monte Carlo simulation.
Simulations are performed for $g^2/(4J_{\sf is}K)=1.4$ and error bars are smaller than the point size.
$\mathcal{O}_{\sf loop}$ takes on a finite value for $T<T_1$, showing the build-up of spin correlations on the loops in the spin-lattice liquid phase.
}}
\label{fig:loopOP}
\end{figure}

Monte Carlo simulations show two phase transitions.
At the high-temperature first-order phase transition ($T=T_1$) the lattice displacement parameter, $\delta \alpha$, acquires a finite value.
For  $T<T_1$ the bonds can therefore be divided into long, medium and short lengths with differing spin interactions, and it becomes meaningful to discuss the system in terms of loops of alternating long and short bonds.
A non-local loop-spin order parameter can be constructed as the product of Ising spins around the loop projected onto its ground state value and given by,
\begin{align}
\mathcal{O}_{\sf loop} = \sum_{\sf loops} \left|  \sum_{i\in {\sf loop}} \sigma_i \sigma_i^{\sf loop,gs}  \right|,
\label{eq:Oloop}
\end{align}
where $\mathcal{O}_{\sf loop}=0 $ for loops with uncorrelated spins and $\mathcal{O}_{\sf loop}=1 $ in the ground state.
The behaviour of this order parameter in Monte Carlo simulations is shown in Fig.~\ref{fig:loopOP}, and it can be seen that it jumps from $\mathcal{O}_{\sf loop} \approx 0$ for $T>T_1$ to a finite value for $T<T_1$.
Further reducing the temperature builds up spin correlations on the loops and therefore increases the value of $\mathcal{O}_{\sf loop}$.

\begin{figure}[h]
\centering
\includegraphics[width=0.4\textwidth]{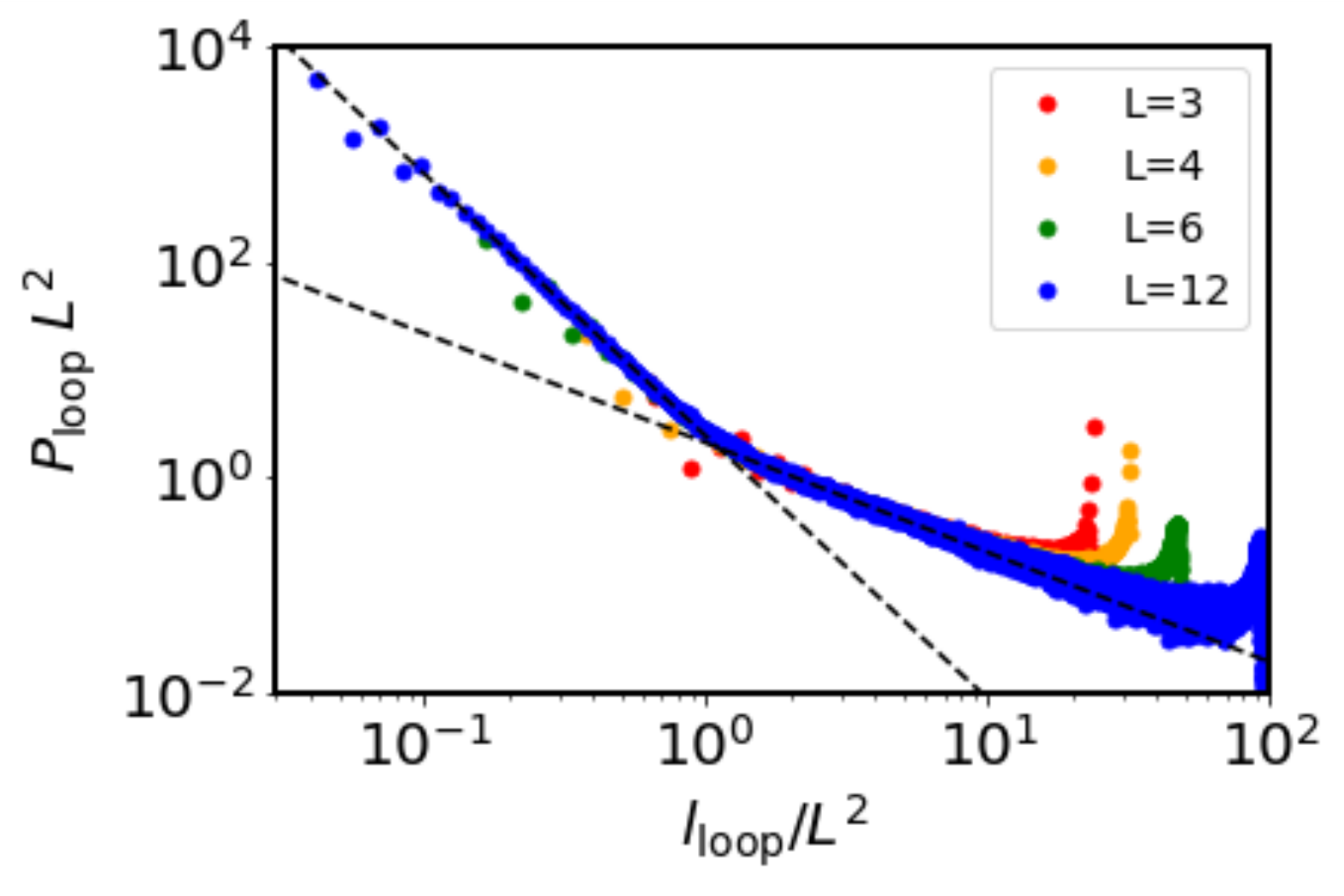}
\caption{\footnotesize{
Loop length distribution in the bulk of the spin-lattice-liquid phase at $T=1.2J_{\sf is}$ for $g^2/(4J_{\sf is}K)=1.4$.
$P_{\sf loop}$ denotes the probablity of a loop having a given length.
There is a crossover between two different power-law distributions at $l_{\sf loop}\approx L^2$.
For $l_{\sf loop}<L^2$ we find $P_{\sf loop}\propto l_{\sf loop}^{-2.44}$ and for $l_{\sf loop}>L^2$ we find $P_{\sf loop}\propto l_{\sf loop}^{-1.02}$.
Within the accuracy of the simulations, these power laws are equivalent to those determined for a 2-in-2-out manifold of states without additional interactions beyond those enforcing the ice rules \cite{jaubert11}, showing that in the spin-lattice liquid phase the spin-mediated interactions between lattice displacements don't significantly affect the lattice-displacement distribution.
}}
\label{fig:loopdist}
\end{figure}

The distribution of loop lengths can be measured by Monte Carlo simulations.
In the bulk of the intermediate phase ($T_2<T<T_1$) we find a power-law distribution of loop lengths, with no sign of a symmetry breaking between loops of length $l_{\sf loop} = 4+4m$ and $l_{\sf loop} = 2+4m$.
This is shown in Fig.~\ref{fig:loopdist}, where it can be seen that there is a crossover between different power laws at $l_{\sf loop} = L^2$.
The crossover is the effect of performing simulations on finite size clusters, where loops can be divided into those that wind or don't wind the system \cite{jaubert11}.
The power-law distribution matches well to that already determined for a 2-in-2-out manifold of states without further interactions \cite{jaubert11}.
As such, the fact that lattice displacements are effectively coupled via the spin degrees of freedom, doesn't restrict their exploration of the full 2-in-2-out manifold in the bulk of the spin-lattice-liquid phase. 

\begin{figure}[h]
\centering
\includegraphics[width=0.35\textwidth]{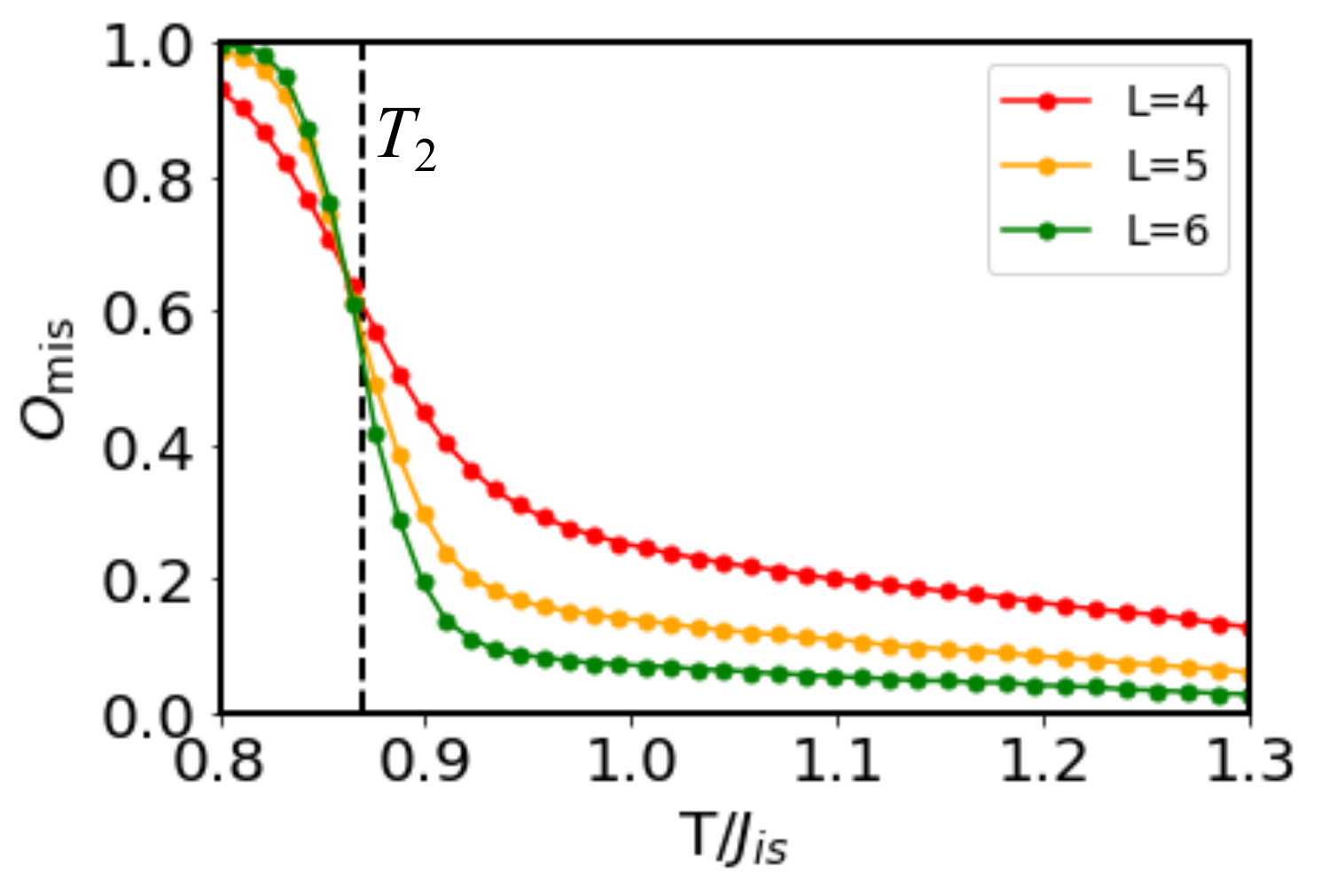}
\caption{\footnotesize{
Loop-length mismatch order parameter, $\mathcal{O}_{\sf mis}$ [Eq.~\ref{eq:Omis}], as measured by Monte Carlo simulation.
Simulations are performed for $g^2/(4J_{\sf is}K)=2$.
$\mathcal{O}_{\sf mis}$ rapidly increases at $T=T_2$, showing that loops of length $l_{\sf loop} = 2+4m$ are frozen out of the system in favour of loops of length $l_{\sf loop} = 4+4m$.
}}
\label{fig:mismatch}
\end{figure}

At lower temperature, Monte Carlo simulations show that there is a phase transition at $T=T_2$ at which the system begins to favour loops of length $l_{\sf loop} = 4+4m$ over those with $l_{\sf loop} = 2+4m$.
This can be seen from studying the distribution of loop lengths, and a simple way of doing this is by considering,
\begin{align}
\mathcal{O}_{\sf mis} = \frac{N_{4+4m} - N_{2+4m}}{N},
\label{eq:Omis}
\end{align}
where $N_{i+4m}$ is the number of sites belonging to loops of lengths $l_{\sf loop} = i+4m$.
$\mathcal{O}_{\sf mis}$ measures the mismatch in the number of sites belonging to the two loop-length classes, and simulation results for its behaviour can be seen in Fig.~\ref{fig:mismatch}.
It takes a relatively low value for $T>T_2$ before rapidly increasing at $T_2$ and saturating at $\mathcal{O}_{\sf mis}=1$ approaching $T=0$.
It can be seen in Fig.~\ref{fig:mismatch} that increasing the system size makes the transition sharper, but the reliability of finite size scaling analysis is hampered by equilibration difficulties.


\subsection{Structure factor}


For comparison with neutron scattering studies it is useful to calculate the structure factor. Since neutrons scatter both off nuclei and from magnetic moments, we calculate both the spin and lattice structure factors.
\begin{figure}[h]
\centering
\includegraphics[width=0.7\textwidth]{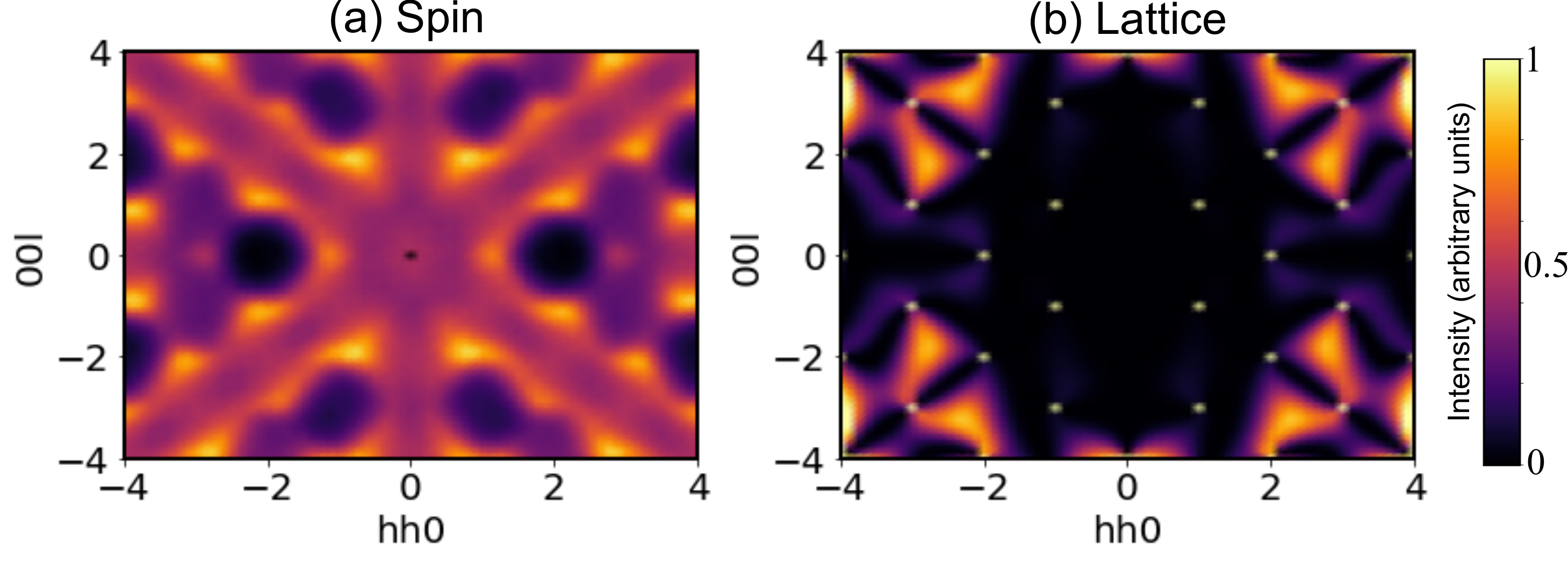}
\caption{\footnotesize{
Spin and lattice structure factors $\mathcal{S}_{\sf sp}({\bf q})$ [Eq.~\ref{eq:Sqspin}] and $\mathcal{S}_{\sf lat}({\bf q})$ [Eq.~\ref{eq:Sqlat}], as calculated by Monte Carlo simulations.
The structure factors are shown deep in the spin-lattice liquid state, with parameters $g^2/(4J_{\sf is}K)=2$, $T=2J_{\sf is}$ and $L=6$.
An $hhl$ cut it taken in reciprocal space.
(a) The spins are correlated but disordered and $\mathcal{S}_{\sf sp}({\bf q})$ therefore shows a diffuse spectrum with well-developed structure but no Bragg peaks. 
(b) The Mo and O nuclei fluctuate around their average positions and therefore $\mathcal{S}_{\sf lat}({\bf q})$ shows a diffuse background superposed with Bragg peaks (shown as white dots, not to scale).
}}
\label{fig:SqSLL}
\end{figure}

The spin structure factor is defined by,
\begin{align}
\mathcal{S}_{\sf sp}({\bf q}) = \sum_{\alpha}\langle S^\alpha({\bf q}) S^\alpha(-{\bf q}) \rangle ,
\label{eq:Sqspin}
\end{align}
where $ S^\alpha({\bf r})$ denotes the $\alpha$th component of spin, measured in the crystallographic coordinates and at position ${\bf r}$, and $ S^\alpha({\bf q})$ is its Fourier transform.
It should be remembered that in the spin-lattice liquid region, the lattice positions of the spins are not fixed but dynamical, and it is necessary to take this into account.
Fig.~\ref{fig:SqSLL}{\red a} shows a cut through the spin structure factor in the spin-lattice liquid state, and displays structure arising from spin correlations.
The cut taken is the $hhl$ plane, where $hkl$ are used to label $(q_{\sf a}/2\pi,q_{\sf b}/2\pi,q_{\sf c}/2\pi)$.

Due to the disordered lattice displacements in the spin-lattice liquid state, nuclear scattering also contributes to the  diffuse scattering \cite{paddison18}.
The total nuclear scattering from Mo ions and their surrounding O octahedra is given by,
\begin{align}
\mathcal{S}_{\sf lat}({\bf q}) = \langle \left|  b_{\sf Mo} e^{i{\bf q}.{\bf r}_{\sf Mo}} +b_{\sf O} e^{i{\bf q}.{\bf r}_{\sf O}} \right|^2 \rangle,
\label{eq:Sqlat}
\end{align}
where $b_{\sf Mo}=6.715$fm and $b_{\sf O}=5.803$fm are the neutron scattering lengths of Mo and O and ${\bf r}_{\sf Mo}$ and ${\bf r}_{\sf O}$ denote the positions of the Mo and O nuclei.
An example structure factor in the spin-lattice liquid state is shown in Fig.~\ref{fig:SqSLL}{\red b} and includes both Bragg peaks arising from the average nuclear positions, as well as diffuse scattering arising from the correlated but disordered displacements.

The total neutron scattering signal will have contributions from both the spin and lattice structure factors, with the relative weight dependent on material parameters.


\end{document}